\documentclass{aa}

\usepackage{graphicx}
\usepackage{txfonts}
\usepackage{natbib}
\usepackage{amsmath}
\usepackage{color}
\usepackage{longtable}
\usepackage{lscape}
\usepackage{enumerate}
\usepackage{url}
\usepackage{mathtools}

\def\msol{\hbox{\kern 0.20em $M_\odot$}}
\def\lsol{\hbox{\kern 0.20em $L_\odot$}}
\def\rsol{\hbox{\kern 0.20em $R_\odot$}}
\def\sr{\hbox{\kern 0.20em sr}}
\def\srmu{\hbox{\kern 0.20em sr$^{-1}$}}

\def\g{\hbox{\kern 0.20em g}}
\def\gmu{\hbox{\kern 0.20em g$^{-1}$}}
\def\kg{\hbox{\kern 0.20em kg}}
\def\pc{\hbox{\kern 0.20em pc}}

\def\mum{\hbox{\kern 0.20em $\mu$m}}
\def\mumd{\hbox{\kern 0.20em $\mu$m$^{-2}$}}
\def\cm{\hbox{\kern 0.20em cm}}
\def\m{\hbox{\kern 0.20em m}}
\def\km{\hbox{\kern 0.20em km}}
\def\nm{\hbox{\kern 0.20em nm}}

\def\s{\hbox{\kern 0.20em s}}
\def\h{\hbox{\kern 0.20em h}}
\def\sec{\hbox{\kern 0.20em sec}}
\def\min{\hbox {\kern 0.20em min}}
\def\smu{\hbox{\kern 0.20em s$^{-1}$}}
\def\smd{\hbox{\kern 0.20em s$^{-2}$}}
\def\an{\hbox{\kern 0.20em an}}
\def\anmu{\hbox{\kern 0.20em an$^{-1}$}}
\def\deg{\hbox{\kern 0.20em $^{\rm o}$}}
\def\yr{\hbox{\kern 0.20em yr}}
\def\yrmu{\hbox{\kern 0.20em yr$^{-1}$}}
\def\Myr{\hbox{\kern 0.20em Myr}}
\def\Mymu{\hbox{\kern 0.20em Myr$^{-1}$}}
\def\K{\hbox{\kern 0.20em K}}
\def\pcmu{\hbox{\kern 0.20em pc$^{-1}$}}
\def\pcmd{\hbox{\kern 0.20em pc$^{-2}$}}
\def\pcmt{\hbox{\kern 0.20em pc$^{-3}$}}
\def\kms{\hbox{\kern 0.20em km\kern 0.20em s$^{-1}$}}
\def\kmpd{\hbox{\kern 0.20em km$^{2}$}}
\def\kpc{\hbox{\kern 0.20em kpc}}
\def\cms{\hbox{\kern 0.20em cm\kern 0.20em s$^{-1}$}}
\def\erg{\hbox{\kern 0.20em erg}}
\def\ergs{\hbox{\kern 0.20em erg}}
\def\cmpd{\hbox{\kern 0.20em cm$^2$}}
\def\cmmd{\hbox{\kern 0.20em cm$^{-2}$}}
\def\cmms{\hbox{\kern 0.20em cm$^{-6}$}}
\def\cmpt{\hbox{\kern 0.20em cm$^3$}}
\def\cmmt{\hbox{\kern 0.20em cm$^{-3}$}}
\def\mpd{\hbox{\kern 0.20em m$^2$}}
\def\mmd{\hbox{\kern 0.20em m$^{-2}$}}
\def\mpt{\hbox{\kern 0.20em m$^3$}}
\def\mmt{\hbox{\kern 0.20em m$^{-3}$}}
\def\mujy{\hbox{\kern 0.20em $\mu$Jy}}
\def\mjy{\hbox{\kern 0.20em mJy}}
\def\Mj{\hbox{\kern 0.20em MJy}}
\def\jy{\hbox{\kern 0.20em Jy}}
\def\ghz{\hbox{\kern 0.20em GHz}}
\def\srmd{\hbox{\kern 0.20em sr$^{-1}$}}

\def \mum{$\mu$m}
\def\G{\hbox{\kern 0.20em G}}

\def\h13cop{\hbox{H$^{13}$CO$^{+}$}}

\def\h2o{\hbox{H$_2$O}}

\begin{document}

   \title{Deuterated methanol toward NGC~7538-IRS1\thanks{\bf{This paper is dedicated to the memory of our dear colleague, Dr. Li-Hong~Xu, who recently passed away.}}}


   \author{J. Ospina-Zamudio
          \inst{1}
          \and
          C. Favre\inst{{1,2}}
          \and
          M. Kounkel\inst{3} 
          \and
          L-H. Xu\inst{{4,\star}}
          \and
          J. Neill\inst{5}
          \and
          B. Lefloch\inst{1}
          \and
          A. Faure\inst{1}          
           \and
          E. Bergin\inst{5}
          \and
          D. Fedele\inst{2} 
          \and
          L. Hartmann\inst{5}
                         }

   \institute{Univ. Grenoble Alpes, CNRS, IPAG, F-38000 Grenoble, France\\
              \email{juan-david.ospina-zamudio@univ-grenoble-alpes.fr}
         \and
             INAF-Osservatorio Astrofisico di Arcetri, Largo E. Fermi 5, I-50125, Florence, Italy
         \and
             Department of Physics and Astronomy, Western Washington University, 516 High St, Bellingham, WA 98225       
         \and
            Centre for Laser, Atomic, and Molecular Sciences (CLAMS), Department of Physics, University of New Brunswick, PO Box 5050, Saint John, NB, Canada E2L 4L5
         \and
             Department of Astronomy, University of Michigan, 1085 South University Avenue, Ann Arbor, Michigan 48109, USA
             }



  \abstract
   {We investigate the deuteration of methanol towards the high-mass star forming region {NGC~7538-IRS1}. We have carried out a multi-transition study of CH$_3$OH, $^{13}$CH$_3$OH and of the deuterated {flavors}, CH$_2$DOH and CH$_3$OD, between 1.0--1.4~mm with the IRAM-30~m antenna. {In total, 34} $^{13}$CH$_3$OH, 13 CH$_2$DOH lines and 20 CH$_3$OD lines spanning a wide range of {upper-state energies (E$_{up}$)} were detected. From the detected transitions, we estimate that the measured D/H does not exceed 1$\%$, with a measured {CH$_2$DOH/CH$_3$OH and CH$_3$OD/CH$_3$OH} of about 
{(32$\pm$8)$\times$10$^{-4}$ and (10$\pm$4)$\times$10$^{-4}$,} respectively. This finding is consistent with the hypothesis of a short-time scale formation during the pre-stellar phase. We find a relative abundance ratio CH$_2$DOH/CH$_3$OD of 3.2 $\pm$ 1.5. This result is consistent with a statistical deuteration. We cannot exclude H/D exchanges between water and methanol if water deuteration is of the order 0.1$\%$, as suggested by recent Herschel observations. 
}

   \maketitle
%

%
\section{Introduction}
\label{sec:introduction}
Observational studies of deuterated molecules are powerful {ways} to probe the chemical and physical evolution of star-forming regions. Indeed, many of the organic species that are abundant constituents of molecular clouds are synthesized in the cold prestellar phase
\citep[see][for a review]{Caselli:2012}. At low temperatures, the  {difference in zero-point energy between deuterated molecules and their hydrogenated counterparts \citep[about 1000~K for methanol and its singly deuterated flavors,][]{Nandi:2018}}  makes it possible for deuterated species to be formed with significantly higher relative abundances than the elemental D/H ratio {\citep[$\sim$10$^{-5}$, see e.g.][]{Ceccarelli:2007,Caselli:2012,Ceccarelli:2014a}.} These enhanced abundance ratios can be preserved as the protostar heats the gas to temperatures large enough ($\ge$100~K) to evaporate the ice mantles. Thus, studies of molecular D/H ratios can provide strong insight into the physical history of star-forming regions along with information on the chemical routes through which the  molecular content is formed.
Deuterium fractionation in low-mass star-forming regions has been the subject of significant study {\citep[e.g., see][]{Roberts:2002,Parise:2006,Ratajczak:2011,Jorgensen:2018}}. These sources have shown remarkably high enhancements of deuterated molecules, including even doubly and triply deuterated species (e.g. ND$_3$, CD$_3$OH), which, in some cases, occur at abundances 12-13 orders of magnitude higher than elemental abundance would suggest \citep{Lis:2002,Parise:2002,Parise:2004}. Regarding high-mass star forming regions, only a few observations of deuterated molecules have been performed so far \citep[e.g.][]{Jacq:1993,Ratajczak:2011,Peng:2012,Neill:2013,Neill:2013a} the best studied high-mass source of deuterated molecules being the Orion KL nebula. Low levels of fractionation are typically observed, consistent with molecular formation at higher temperatures; alternatively, warm gas-phase chemistry could alter the D/H ratio in high-mass regions during the post-evaporative phase. Interestingly enough, from chemical modelling using experimental kinetic data, \citet{Faure:2015} suggested that the D/H ratio measured in Orion-KL might not be representative of the original mantles due to deuterium exchanges between water and methanol in interstellar ices during the warm up phase. The study of the abundance ratio of {the deuterated isotopologues methanol (CH$_3$OD/CH$_2$DOH)} could give access to the initial deuteration of water ices before the warm up phase sets in. Nonetheless, taking into account the limited number of observations of deuterated species in high-mass {star-forming} regions, the extent to which the D/H ratios observed toward Orion KL are representative of high-mass sources chemistry is uncertain.
\citet{Bogelund:2018} have recently reported low deuteration levels of methanol towards the HMSFR NGC6334~I, with typical values in the range of 0.01--1$\%$ for both CH$_2$DOH and CH$_3$OD. Large uncertainties on the column densities, of a factor of 4 to 10 depending on the species remain however, which prevent any robust conclusions on the CH$_2$DOH/CH$_3$OD ratio and the initial ice composition. Further observations of a sample of high-mass sources are thus required for comparison between sources and for improving our understanding of the chemistry that creates organic molecules with significant deuterium fractionation (in particular, the roles of grain-surface and gas-phase reactions).

In this study, we investigate the D/H ratio  for methanol (CH$_3$OH, CH$_2$DOH and CH$_3$OD) towards the high-mass forming region {NGC~7538-IRS1} \citep[$L=1.3 \times 10^5~L_{\odot}$, $d=2.8$~kpc, $M\sim30~M_{\odot}$, see][]{Bisschop:2007,Beuther:2012}, which is known to harbour high abundances of organic molecules, including methanol \citep[e.g][]{Bisschop:2007,Wirstrom:2011}. Results are given in Section~\ref{sec:results} and discussed in Section~\ref{sec:discussion}.

%
\section{Observations and data reduction}
\label{sec:observations}
%
The data were acquired with the IRAM 30--m telescope on 2013 December 5, 6 {and 10} towards {NGC~7538-IRS1} with a single pointing ($\rm\alpha_{J2000}$=23$\rm^{h}13^{m}45\fs5$, $\rm\delta_{J2000}$= $+$61$\degr28\arcmin12\farcs$0). The $v_{LSR}$ was -57~km~s$^{-1}$.
The observations were taken in position switching mode, using a reference for the OFF position at $[-600 \arcsec,0 \arcsec]$.
The EMIR receiver at 1~mm was used in connection with {the Fourier transform spectrometer (FTS) as backend, providing  a spectral resolution 195~kHz. The spectral resolution was subsequently degraded to 780 KHz, which corresponds to about 1~km~s$^{-1}$, in order to improve the signal-to-noise ratio of the data.}
The following frequency ranges were covered: 212.6--220.4~GHz, 228.3--236.0~GHz, 243.5--251.3~GHz, 251.5--259.3~GHz, 259.3--267.0~GHz and 267.2--274.9~GHz. The half-power beam sizes are 10$^{\prime\prime}$ and 12$^{\prime\prime}$ for observations at 250~GHz and 212~GHz, respectively.

The data were reduced using the CLASS90 package from the GILDAS software\footnote{ http://www.iram.fr/IRAMFR/GILDAS/}.
The spectra are reported in this study in main beam temperature units, T$\rm_{MB}$, given by $T\rm_{MB}= {\eta\rm_{f} \over \eta\rm_{MB}} \times T\rm_{A}^{*}$, where T$\rm_{A}^{*}$ is the antenna temperature, $\eta\rm_{f}$ the forward efficiency ($\eta\rm_{f}$ = 94, 92 and 87 at 210, 230 and 274~GHz, respectively)
and $\eta\rm_{MB}$ the main beam efficiency ($\eta\rm_{MB}$= 63, 59, 49 at 210, 230 and 274~GHz, respectively\footnote{see http://www.iram.es/IRAMES/mainWiki/Iram30mEfficiencies}).
Finally, the data presented some spurs along with standing waves, which have been removed during the data reduction.
 In this paper, we focus on methanol and its deuterated flavors.
 
  The spectroscopic parameters of the targeted lines are given in Appendix~\ref{appA} (see Tables~\ref{taba1} to \ref{taba4}). More specifically, we used the spectroscopic data parameters\footnote{The spectroscopic data parameters are available at the Cologne Database for Molecular Spectroscopy catalog \citep[CDMS][]{Muller:2005} and/or at the JPL catalog \citep{Pickett:1998}.}
from {\citet{Xu:1995,Xu:1995a}, \citet{Xu:1997}, \citet{Xu:2008} and \citet{Muller:2004a}} for CH$_3$OH. The parameters for CH$_2$DOH come from \citet{Pearson:2012} and that for $^{13}$CH$_3$OH are from \citet{Xu:1997,Xu:1996,Xu:2008,Xu:2014}. Regarding CH$_3$OD, we used the same unpublished data as {\citet{Neill:2013a}} used for analysing the HIFI/Herschel observations of Orion-KL. Nonetheless, the CH$_3$OD partition functions we use in the present study \citep[see also,][]{Walsh:2000} are given in Appendix~\ref{appB}. Finally, collisional rates are assumed to be identical to those of CH$_3$OH and taken from \citet{Rabli:2010}. {This point is addressed in more detail in Section 3.3}

%
\section{Results and analysis}
\label{sec:results}

\begin{figure}
\includegraphics[width=\columnwidth]{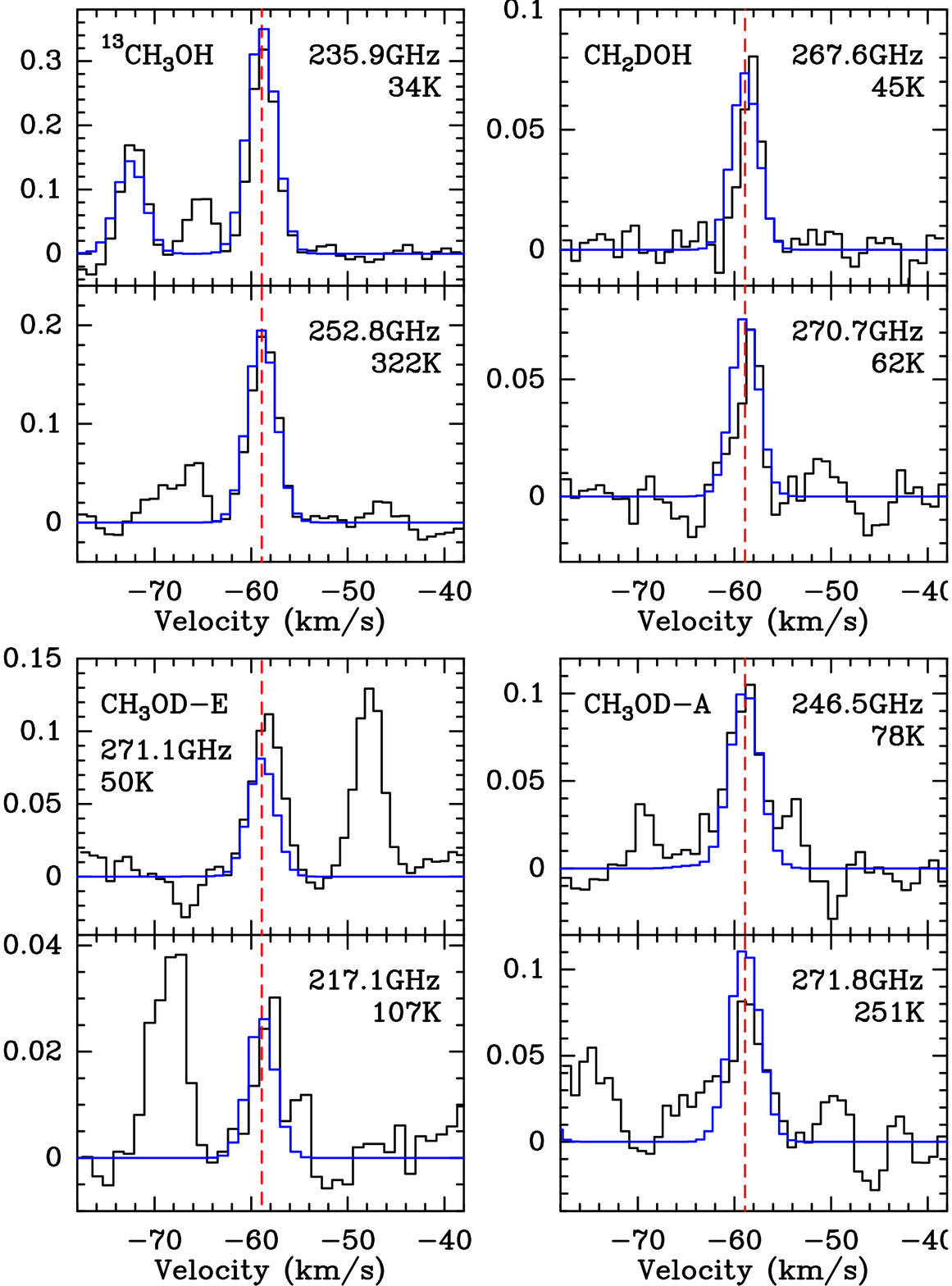}
\includegraphics[width=\columnwidth/2]{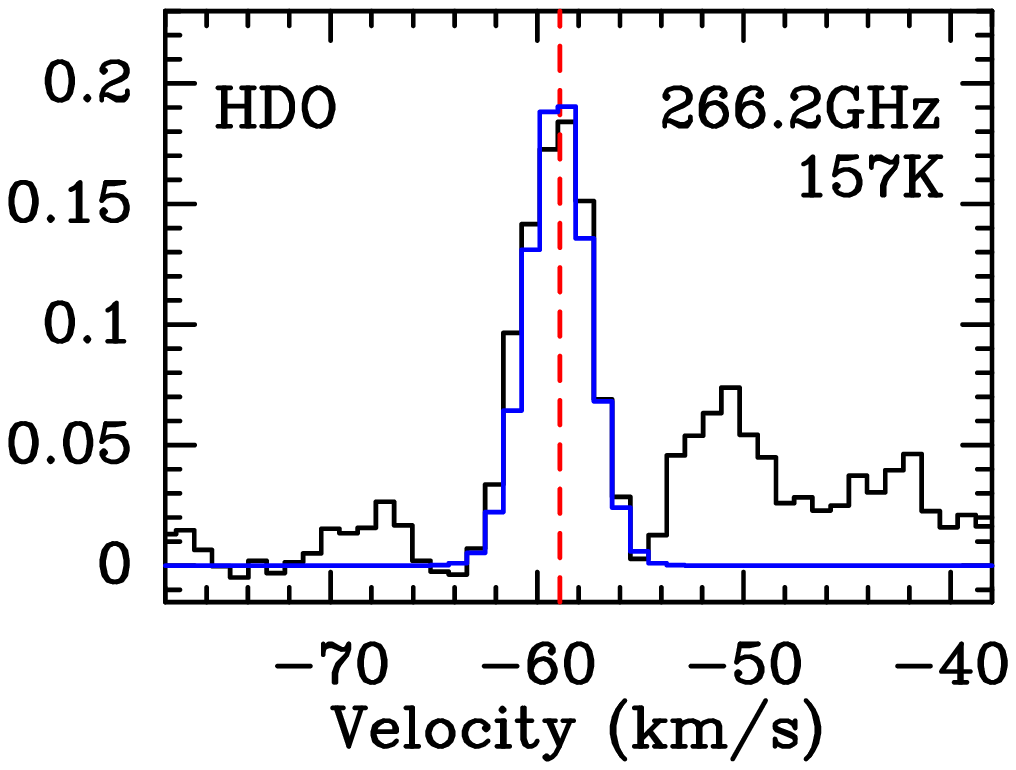}
\caption{\label{fig:spectra}Montage of detected transitions of methanol $^{13}$C isotopologue and deuterated forms along with that of the detected HDO line. Intensities are expressed in unit of $T_{mb}$. {Our LTE modelling is displayed in blue}. The red dashed line marks the peak velocity of methanol transitions $v_{LSR} = -58.9\kms$.}
\end{figure}

\subsection{Detected Lines}

We detect several bright lines covering a wide range of upper energy levels: 34 $^{13}$CH$_3$OH lines with $E_{up}$ spanning from 23 to 397~K; 13 CH$_2$DOH lines with $E_{up}$ spanning from 25 to 94~K; and 20 CH$_3$OD lines (10 lines for both A- and E- forms) with $E_{up}$ spanning from 19 to 271~K. The observational line parameters of the clearly detected transitions are summarised in Tables~\ref{taba1} to \ref{taba4}).

\begin{figure*}
\center
\includegraphics[width=2\columnwidth]{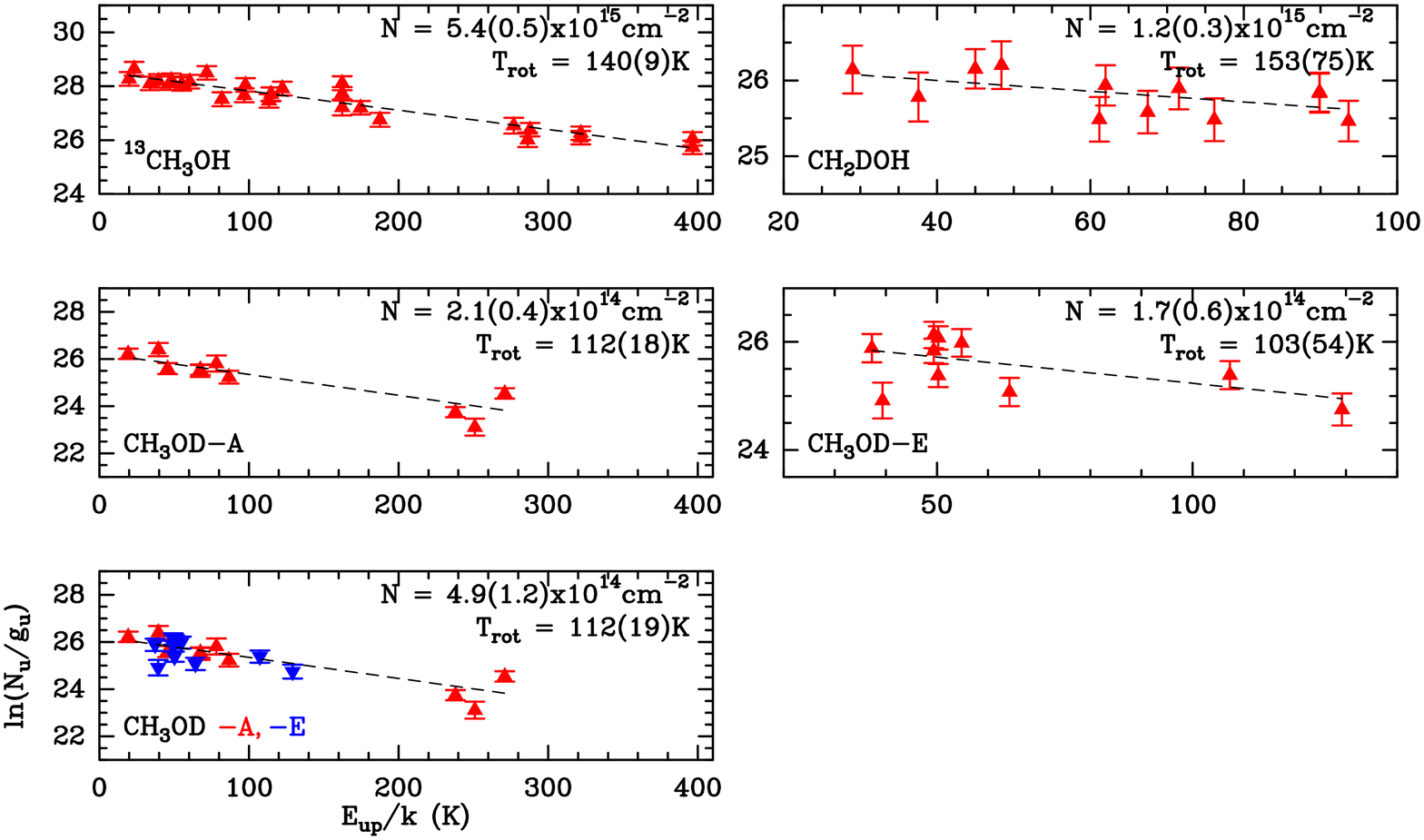}
\caption{\label{fig:popdiag}Population diagram analysis of the $^{13}$C-methanol and the deuterated forms. A size of 3.8\arcsec\ was adopted for the emitting region. {In the analysis we separate the methanol E-form from the A-form. However, as shown in the {bottom left panel,} their respective SLED are consistent with each-other. {Finally, please note that the y-range varies from plot to plot.}}}
\end{figure*}

The lines profiles are well fitted by a single Gaussian profile with a little scatter on the fitted full-width half-maximum (mean $\Delta v = 3.4 \pm 0.7\kms$) and the peak intensity velocity (mean $v_{LSR}= 58.6 \pm 0.6\kms$). {We note that the scatter on the peak velocity lies within 1 element of spectral resolution (1~km~s$^{-1}$), which makes us confident with the line assignations. In addition, if we take into account the pointing uncertainties (typically 1$\arcsec$--2$\arcsec$), the observed scatter in emission velocity peak is consistent with the large velocity gradient observed within the source by \citet[][and their Fig.~8]{Beuther:2012}.}
Figure~\ref{fig:spectra} show a montage of 2 detected transitions probing different excitation energies of the following species: $^{13}$CH$_3$OH,  CH$_2$DOH, CH$_3$OD (A and E) along with {our  LTE modelling}.
In addition, Fig.~\ref{fig:spectra} also displays the spectrum of the 2$_{2,0}$-3$_{1,3}$ transition of deuterated water, that is also detected in our survey. 
{Finally, Figures~\ref{fig:c1} to \ref{fig:c3} (in Appendix~\ref{appC}) display a montage of all the detected transitions associated with the $^{13}$CH$_3$OH,  CH$_2$DOH, CH$_3$OD species along with our LTE modeling}.

\subsection{LTE analysis}

The physical properties (excitation temperature, column densities) of the methanol species were obtained from a population diagram analysis of their Spectral Line Energy Distribution (SLED). In the present study, we assume a source size of 3.8\arcsec, that corresponds to the ice evaporation region \citep[see][]{Bisschop:2007}. The derived column densities are {therefore determine and given for a source size 3.8\arcsec }. We estimate the line opacities, under Local Thermodynamical Equilibrium (LTE) conditions following \citet[see]{Goldsmith:1999}. We conclude that the $^{13}$CH$_3$OH, CH$_2$DOH and CH$_3$OD emission are optically thin with $\tau \leqslant 0.03$.
{We note that most of the transitions associated with the main methanol isotopologue are optically thick with $\tau \rm{(CH_3OH)} \ge 1$. As a consequence, we cannot determine with an accurate enough precision  the excitation temperature and the column density of the $^{12}$CH$_3$OH. We therefore exclude the analysis of the $^{12}$CH$_3$OH for the present study.}

Figure~\ref{fig:popdiag} shows the population diagrams that are all well fitted by a single rotational temperature. More specifically, the population diagram analysis of $^{13}$CH$_3$OH yields to $N$($^{13}$CH$_3$OH$) = (5.4 \pm 0.5) \times 10^{15}\cmmd$ and $T_{rot} = 140 \pm 9\K$.  Assuming a $^{12}$C/$^{13}$C elemental abundance ratio of 70 for the local ISM \citep{Wilson:1999}, we determine the methanol column density $N$(CH$_3$OH$) = (3.8 \pm 0.4) \times 10^{17}\cmmd$.
Regarding CH$_2$DOH, we estimate a rotational temperature  similar to that of $^{13}$C methanol, $T_{rot} = 153 \pm 75\K$, and a column density $N$(CH$_2$DOH) of (1.2 $\pm$ 0.3) $\times$ 10$^{15}\cmmd$.
 For CH$_3$OD we derive a lower rotational temperature: $T_{rot} = 103 \pm 54\K$ ($112 \pm 18\K$) for both the E- and A- forms.
Nonetheless, within the error bars, the derived rotational temperatures for the deuterated methanol flavors are in agreement with the one of the$^{13}$C isotopologue. The derived column densities of both E- and CH$_3$OD-A are comparable with $N$(CH$_3$OD-E$) = (1.7 \pm 0.6) \times 10^{14}\cmmd$ and $N$(CH$_3$OD-A$) = (2.1 \pm 0.4) \times 10^{14}\cmmd$; which results a total CH$_3$OD column density, $N$(CH$_3$OD$), of (3.8 \pm 1.0) \times 10^{14}\cmmd$. {We note that the computed total CH$_3$OD column density is commensurate within the error bars to that found if we treat both substates (E- and A-) simultaneously (see Fig.~\ref{fig:popdiag}).}

\begin{table}
\caption{Physical properties of methanol isotopologues: rotational temperature and column density. A source size of 3.8" was adopted.}
\center
\begin{tabular}{lrr}

\hline
\hline
Species 		& $T_{rot}$	& $N$			\\
			& (K)		& ($10^{14}\cmmd$)	\\
\hline
$^{13}$CH$_3$OH		& $140\pm9$	& $54\pm5$		\\
CH$_2$DOH		& $153\pm75$	& $12\pm3$		\\
{CH$_3$OD-E}	& $103\pm54$	& $1.7\pm0.6$		\\
{CH$_3$OD-A}		& $112\pm18$	& $2.1\pm0.4$		\\
\hline
\end{tabular}
\end{table}

\subsection{non-LTE analysis}
The present section aims to verify whether the apparent LTE distribution of the CH$_3$OD targeted transitions is consistent with non-LTE conditions along with the derived densities and temperatures. 

For this purpose we have combined the (unpublished) spectroscopic data of CH$_3$OD (energy levels and radiative rates) with the collisional rate coefficients computed by \citet{Rabli:2010} for the rotational excitation of CH$_3$OH by H$_2$. We thus assumed that the -OH H/D substitution has a negligible impact on the scattering dynamics (the change in reduced mass is only 0.2\%). The coupled statistical equilibrium-radiative transfer equations were then solved using the RADEX code \citep{van-der-Tak:2007}. 

\begin{figure}
\includegraphics[angle=270,width=1.\columnwidth, clip, trim =1cm 2.cm 0.cm 5cm]{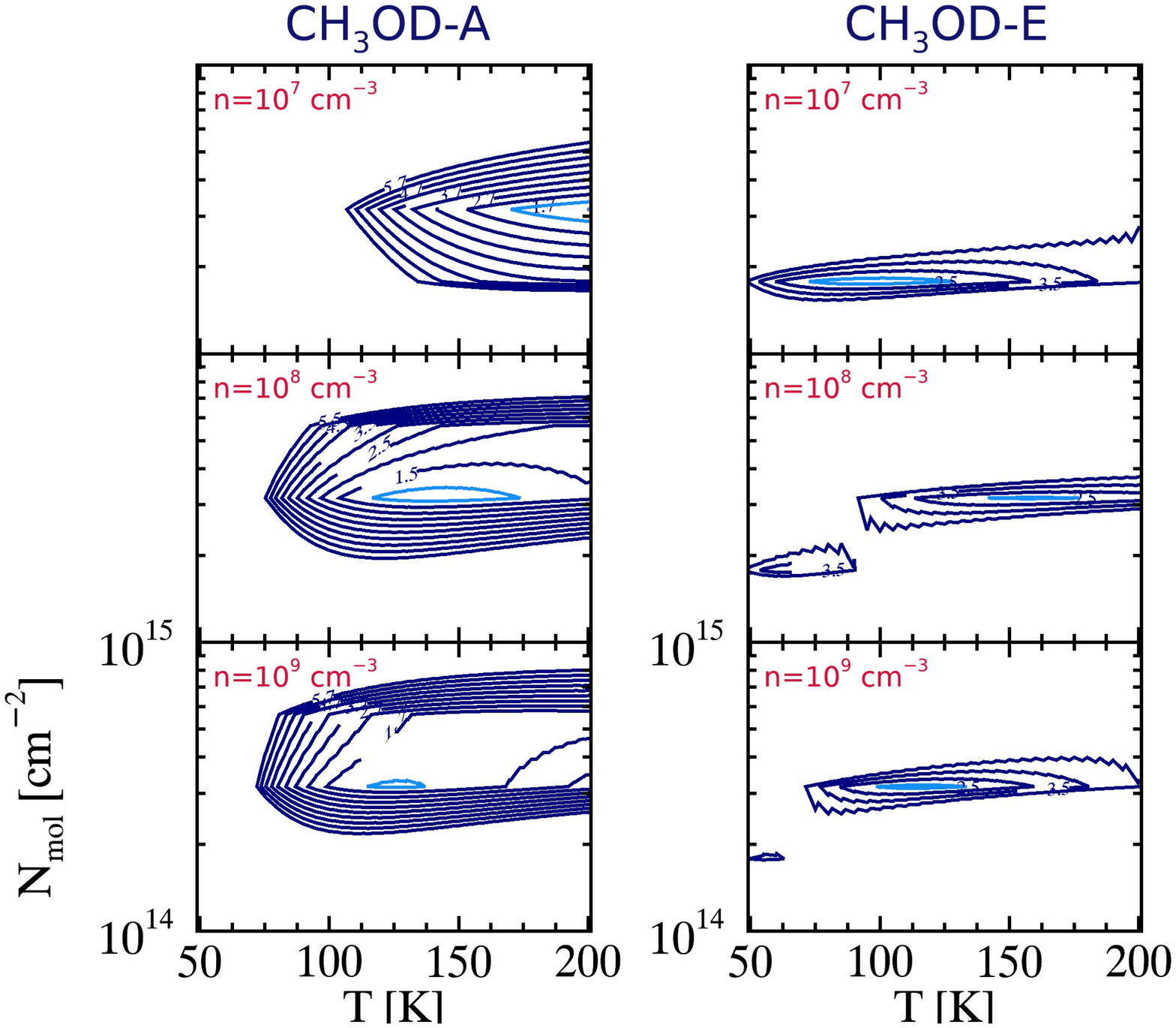}
\caption{\label{fig:radex}{CH$_3$OD-A (left column) and CH$_3$OD-E (right column) $\chi^{2}$ distributions as computed with RADEX for n=10$^{7}$, 10$^{8}$ and 10$^{9}$~cm$^{-3}$. The minimum $\chi^{2}$ is displayed in cyan.}}
\end{figure}

Assuming a source size of 3.8$\arcsec$ and a line-width of 4~km~s$^{-1}$, we ran a grid of models for the following gas densities n=10$^{7}$, 10$^{8}$ and 10$^{9}$~cm$^{-3}$ with the kinetic temperature varying from 20 to 200~K and the CH$_3$OD column density lying between 1$\times$10$^{10}$ and 1$\times$10$^{19}$~cm$^{-2}$. Then, we compare the observed and modelled LGV {CH$_3$OD-A} and {CH$_3$OD-E} integrated fluxes by the mean of the $\chi^{2}$. The computed results and minimized $\chi^{2}$ distributions are shown in Figure~\ref{fig:radex}.

There is no common solution for both E- and A-species with a density of $10^7$. A common region of minimum $\chi^{2}$  in  the parameter space (N,T) is found for a density in the range $10^8\cmmt$-$10^9\cmmt$. The latter is commensurate with the one derived by \citet{Beuther:2012} from PdBI observations as well as that derived by \citet{Bisschop:2007} from JCMT observations. 
Our best fits are obtained for a CH$_3$OD column density of about $\simeq 3\times 10^{14}\cmmd$, a temperature 100--$150\K$, both consistent with our LTE analysis. 

%
\section{Discussion}
\label{sec:discussion}
\subsection{D/H in methanol}

\begin{table}
\caption{Deuteration ratios measured towards {NGC~7538-IRS1}.}
\center
\begin{tabular}{lr}
\hline
\hline
Species 		& Ratios	\\
\hline
CH$_2$DOH/CH$_3$OH	& $(3.2\pm0.8)\times10^{-3}$	\\
CH$_3$OD/CH$_3$OH	& $(1.0\pm0.4)\times10^{-3}$	\\
CH$_2$DOH/CH$_3$OD	& $3.2\pm1.5$			\\
\hline
\end{tabular}
\end{table}

As can be seen in Table~2, the fractionation degree of singly deuterated methanol is low in {NGC~7538-IRS1}, with relative abundance values of 0.1$\%$ (taking into account the statistical ratio for the -CH$_3$ functional group). Such values are similar to those reported towards
other HMSRs like Orion KL.  Thanks to the numerous detected lines for both CH$_2$DOH and CH$_3$OD, the relative abundance between the two deuterated flavors is determined with a good accuracy, and is found to equal to 3.2 $\pm$1.5.

\subsection{Water and HDO}
Water does not exhibit the same level of fractionation as methanol in {low-mass star-forming regions, with a HDO/H$_2$O lying} in the range 0.01-0.07 \citep{Liu:2011,Parise:2005,Parise:2006,Ratajczak:2011,Coutens:2012,Faure:2015}. However, in Orion-KL water and methanol seem to be fractionated to a similar extent \citep{Neill:2013a}. This is likely the result of recent desorption from ice mantles \citep[see further details on thermal H/D exchanges between water and methanol during the warm-up phase in][]{Faure:2015}.

Although only the HDO (2$_{2,0}$-3$_{1,3}$) transition is detected in our data, as shown in Fig.~1, we can roughly estimate the D/H ratio for water combining our result to that of different studies.
Indeed, combining the observed HDO integrated flux ($\sim$ 0.8$\pm$0.1~K~km~s$^{-1}$) to the ones measured by \citet{Jacq:1990} for the (3$_{1,2}$-2$_{2,1}$) and (2$_{1,1}$-2$_{1,2}$) transitions and, assuming a source size of 3.8$\arcsec$, we derived a HDO column density of about 1.6$\times$10$^{15}$~cm~$^{-2}$ and a rotational temperature of $\sim$ 130~K.

From recent Herschel/HIFI observations of water in {NGC~7538-IRS1}, \citet{Herpin:2017} have derived a water abundance,  $\chi_{\rm{(H{_2}O)}}$, of $8\times 10^{-6}$. Assuming a n$\rm_{H{_2}}$ of about 10$^{8}$~cm$^{-3}$ (see Section~3.3), we estimate a HDO/H$_2$O ratio lying in the range $10^{-4}$-$10^{-3}$ according to the source size taken into account \citep[see also][]{van-der-Tak:2006}.

Incidentally, we note that \citet{Herpin:2017} also derived a H$_2$O abundance of $5\times 10^{-5}$ for the very inner part of the hot core, as probed by a THz line observed with SOFIA. 

These findings show that the water emission source size gives strong constraints on the deuteration ratio. Therefore, further {interferometric} observations of water and its isotopologues are clearly needed to investigate in more detail water fractionation in this source.

\subsection{Comparison with other sources}
\begin{figure}
\center
\includegraphics[width=0.95\columnwidth, clip, trim =2.5cm 17.4cm 3.9cm 2.5cm]{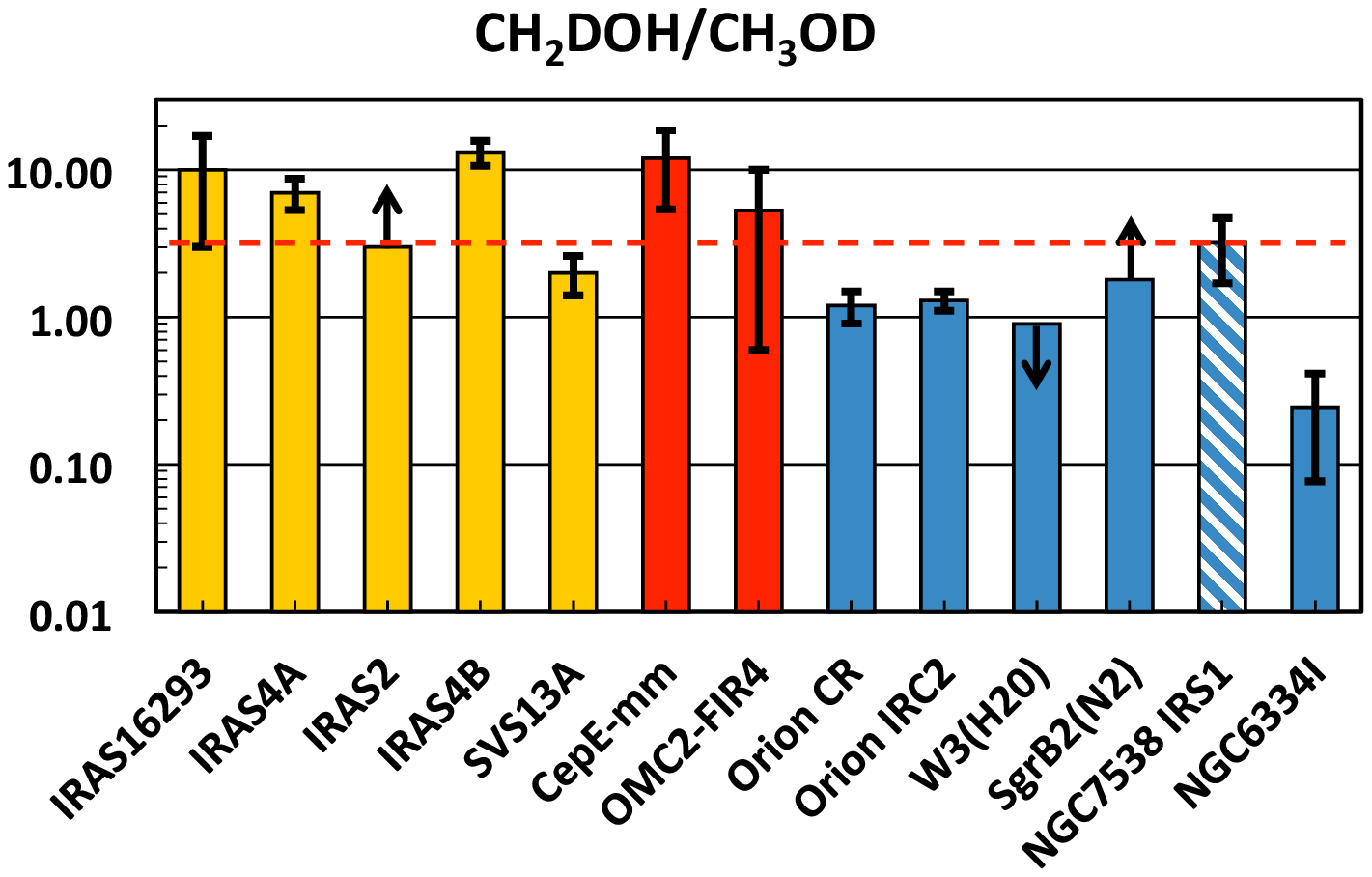}
\caption{\label{fig:deut__ratio_sources}CH$_2$DOH/CH$_3$OD ratios observed towards low (yellow), intermediate (red) and high mass (blue) star forming regions: IRAS16293-2422 \citep{Parise:2002,Jorgensen:2018}, IRAS4A, IRAS2, IRAS4B \citep{Parise:2006}, SVS13A \citep{Bianchi:2017}, CepE-mm \citep{Ratajczak:2011}, OMC2-FIR4 \citep{Ratajczak:2011}, Orion CR \citep{Neill:2013a}, Orion Irc2, W3(H2O), SgrB2(N2) \citep{Belloche:2016}, {NGC~7538-IRS1} (this work; blue and white stripes) and NGC6334I \citep{Bogelund:2018}. We note that values derived by \citet{Parise:2006} and \citet{Ratajczak:2011} were divided by a factor of 2 and 1.5, respectively, due to a spectroscopic issue \citep[see][]{Belloche:2016}. The dashed red line shows the statistical factor CH$_2$DOH/CH$_3$OD = 3.}
\end{figure}

We have reported in Fig.~\ref{fig:deut__ratio_sources} the relative CH$_2$DOH/CH$_3$OD abundance ratio measured towards a large sample of  star forming regions, from low- to high-mass, ordered from left to right by increasing source luminosity. 
It is immediately apparent that there is a 2 orders of magnitude variation between the sources. More specifically, 
in low-mass star forming regions, methanol deuteration strongly favors the methyl group (CH$_2$DOH) by far more than the statistical factor (3), whereas in high-mass the opposite trend is observed.

\subsection{Modelling}

The value of the abundance ratio CH$_2$DOH/CH$_3$OD toward NGC~7538-IRS1 is $3.2 \pm 1.5$, in good agreement with the value of 3 predicted by grain chemistry models \citep[e.g.][]{Charnley:1997,Osamura:2004}. In these models, the deuterium fractionation of methanol proceeds in the ice during the early cold prestellar phase through the statistical addition of H and D atoms on CO molecules and the ratio s-CH$_2$DOH/s-CH$_3$OD is equal to the statistical value of 3. More recent and sophisticated models also predicts a ratio close to 3 \citep[e.g.][]{Bogelund:2018}. These models however neglect processes occurring in the subsequent warm-up phase. In particular, \citet{Faure:2015} have shown that the s-CH$_2$DOH/s-CH$_3$OD ratio can change during this phase as a result of H/D exchanges between the hydroxyl (-OH) functional groups of methanol and water \citep[see details in][]{Faure:2015a}. This scenario was successful in explaining both the high value ($>$3) of the CH$_2$DOH/CH$_3$OD abundance ratio towards the low-mass protostar IRAS~16293-2422 and the low-value ($<$3) measured towards Orion (see Fig.~\ref{fig:deut__ratio_sources}).

\begin{figure}
\includegraphics[width=1.1\columnwidth, angle=270,clip,trim =0cm 0.cm 0cm 2.5cm ]{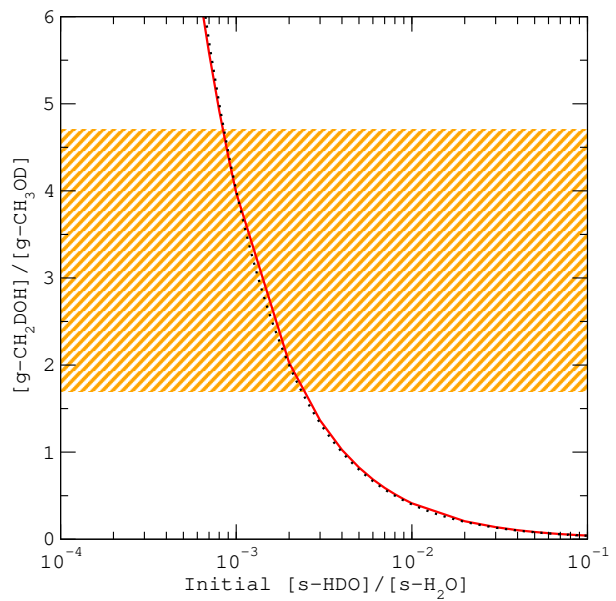}
\caption{\label{fg5}Gas-phase abundance ratio of the deuterated isotopologues, CH$_2$DOH/CH$_3$OD, as function of the initial (cold) water ice deuteration. The dotted line corresponds to the analytic solution CH$_2$DOH/CH$_3$OD=0.004/(HDO/H$_2$O) 
\citep[see][for details]{Faure:2015}.The ratio observed toward NGC~7538-IRS1 is represented by the orange hatched zone.}
\end{figure}

We have adapted the model of \citet{Faure:2015} to the conditions of NGC~7538-IRS1. First, the density was taken as $n_{\rm H}=2\times 10^8$~cm$^{-3}$ and the (equal) gas and dust temperatures as $T=100$~K.  The methanol abundance (by number) relative to water is 4\% \citep{Oberg:2011e,Boogert:2015} and the mean water abundance is $5\times 10^{-5} n_{\rm H}$. The accreting D/H ratio is inferred from the observed CH$_2$DOH/CH$_3$OH ratio (see Table~2) as $\alpha_m=1.1\times10^{-3}$, assuming that the initial (statistical) deuteration of CH$_2$DOH in the ice is conserved during the hot core phase. The post-evaporative gas-phase chemistry is also entirely neglected \citep[full details can be found in ][]{Faure:2015}. The results of the model are plotted in Fig.~\ref{fg5}: it is found that the observed CH$_2$DOH/CH$_3$OD ratio of $3.2 \pm 1.5$ can be reproduced for a solid s-HDO/s-H$_2$O ratio in the range $\sim 8\times 10^{-4}-2\times 10^{-3}$. Thus, while the CH$_2$DOH/CH$_3$OD ratio of $3.2 \pm 1.5$ is consistent with a statistical deuteration, the occurence of H/D exchanges cannot be excluded. In particular, it should be noted that H/D exchanges can explain the range of values depicted in Fig.~\ref{fig:deut__ratio_sources}, that is CH$_2$DOH/CH$_3$OD ratios in the range $\sim 0.1-10$. To our knowledge, this is currently the unique mechanism able to explain the non-statistical deuteration of methanol in both low- and high-mass protostars.

%
\section{Conclusions}
To summarize, we have investigated the deuteration of methanol in the high-mass star forming regions {NGC~7538-IRS1} using IRAM-30m observations. Our study shows that the fractionation degree of deuterated methanol is low in this source. From the numerous ($\ge$10) detected lines, we estimate a CH$_2$DOH/CH$_3$OD relative abundance ratio of $3.2 \pm 1.5$. Although our findings are commensurate with statistical deuteration at the icy surface of grain mantles, we cannot exclude H/D exchanges between water and methanol at the present time. Further observations of water and HDO are required to address this point.
\label{sec:conc}

%
\begin{acknowledgements}
 {We thank the anonymous referee for his very fruitful comments that have strengthened our paper. This work is supported by the French National Research Agency in the
framework of the Investissements d’Avenir program (ANR-15-IDEX-02),
through the funding of the "Origin of Life" project of the Univ. Grenoble-Alpes. } 
CF and DF acknowledge support from the Italian Ministry of Education, Universities and Research, project SIR (RBSI14ZRHR). LHX acknowledges financial support from the Natural Sciences and Engineering Research Council of Canada. {We also  acknowledge the funding from the European Research Council (ERC) under the European Union’s Horizon 2020 research and innovation programme, for the Project "The Dawn of Organic Chemistry" (DOC), grant agreement No 741002. This project was carried out under project number 128--12  with the IRAM 30 m telescope. IRAM is supported by INSU/CNRS (France), MPG (Germany), and IGN (Spain).  }
\end{acknowledgements}

%
%
\bibliographystyle{aa}

%
%
\begin{appendix}

\section{Spectroscopic and Observational line parameters}
\label{appA}
Tables~\ref{taba1} to \ref{taba4} list the spectroscopic and observational line parameters for the observed CH$_3$OH-A, CH$_3$OH-E, $^{13}$CH$_3$OH, CH$_2$DOH, CH$_3$OD-A and CH$_3$OD-E transitions.

\begin{table*}
\caption{Methanol}
\resizebox{1.75\columnwidth}{!}{
\begin{tabular}{lrcrrrrrr}
\hline\hline
Species & Frequency & Quantum numbers & $E_{up}$ & $A_{ul}$ & Flux & V$_{lsr}$ & $\Delta$V & $T_{peak}$ \\
 & (MHz) &  & (K) & (10$^{-5 }$\,s$^{-1}$) & (K\,km\,s$^{-1}$) & (km\,s$^{-1}$) & (km\,s$^{-1}$) & (mK) \\
\hline
CH$_3$OH-E & 213159.150 & 20$_{-4,0}$ -- 19$_{-5,0}$ & 567.0 & 1.6 & 1.6 (0.3) & -59.0 (0.0) & 4.0 (0.1) & 374 (9) \\
 & 213377.528 & 13$_{6,0}$ -- 14$_{5,0}$ & 382.0 & 1.1 & 1.4 (0.3) & -59.0 (0.0) & 3.5 (0.1) & 368 (7) \\
 & 213427.061 & 1$_{1,0}$ -- 0$_{0,0}$ & 15.5 & 3.4 & 4.4 (0.9) & -58.8 (0.0) & 3.5 (0.0) & 1173 (18) \\
 & 217886.504 & 20$_{1,0}$ -- 20$_{0,0}$ & 500.5 & 3.4 & 1.8 (0.4) & -59.2 (0.0) & 3.7 (0.0) & 439 (5) \\
 & 218440.063 & 4$_{2,0}$ -- 3$_{1,0}$ & 37.6 & 4.7 & 10.0 (2.0) & -58.1 (0.0) & 5.3 (0.1) & 1776 (12) \\
 & 219983.675 & 25$_{3,0}$ -- 24$_{4,0}$ & 794.3 & 2.0 & 0.4 (0.1) & -59.0 (0.1) & 3.1 (0.3) & 130 (13) \\
 & 219993.658 & 23$_{5,0}$ -- 22$_{6,0}$ & 768.0 & 1.7 & 0.6 (0.1) & -58.9 (0.1) & 3.8 (0.2) & 143 (8) \\
 & 220078.561 & 8$_{0,0}$ -- 7$_{1,0}$ & 88.7 & 2.5 & 2.9 (0.6) & -58.7 (0.0) & 4.1 (0.1) & 677 (7) \\
 & 229589.056 & 15$_{4,0}$ -- 16$_{3,0}$ & 366.5 & 2.1 & 2.2 (0.4) & -59.1 (0.0) & 3.1 (0.1) & 651 (14) \\
 & 229758.756 & 8$_{-1,0}$ -- 7$_{0,0}$ & 81.2 & 4.2 & 5.0 (1.0) & -58.3 (0.0) & 4.2 (0.0) & 1124 (11) \\
 & 230027.047 & 3$_{-2,0}$ -- 4$_{-1,0}$ & 31.9 & 1.5 & 3.0 (0.6) & -58.6 (0.0) & 4.2 (0.1) & 658 (14) \\
 & 230368.763 & 22$_{4,0}$ -- 21$_{5,0}$ & 674.8 & 2.1 & 1.1 (0.2) & -59.0 (0.2) & 3.9 (0.5) & 274 (62) \\
 & 232624.811 & 15$_{6,1}$ -- 15$_{7,1}$ & 861.9 & 0.7 & 0.3 (0.1) & -58.8 (0.1) & 3.3 (0.3) & 74 (6) \\
 & 232645.103 & 8$_{6,1}$ -- 8$_{7,1}$ & 667.7 & 0.9 & 0.4 (0.1) & -58.8 (0.2) & 4.1 (0.4) & 84 (7) \\
 & 232847.103 & 9$_{6,1}$ -- 9$_{7,1}$ & 688.6 & 1.0 & 0.5 (0.1) & -59.0 (0.1) & 4.2 (0.3) & 106 (6) \\
 & 232945.797 & 10$_{-3,0}$ -- 11$_{-2,0}$ & 182.5 & 2.1 & 3.8 (0.8) & -59.0 (0.0) & 3.6 (0.1) & 999 (15) \\
 & 233011.878 & 10$_{6,1}$ -- 10$_{7,1}$ & 711.7 & 1.1 & 0.3 (0.1) & -58.7 (0.2) & 3.3 (0.4) & 87 (9) \\
 & 233121.162 & 11$_{6,1}$ -- 11$_{7,1}$ & 737.1 & 1.1 & 0.5 (0.1) & -59.2 (0.2) & 4.2 (0.5) & 113 (10) \\
 & 233155.874 & 12$_{6,1}$ -- 12$_{7,1}$ & 764.9 & 1.0 & 0.3 (0.1) & -59.0 (0.3) & 4.0 (0.8) & 68 (10) \\
 & 234698.519 & 5$_{-4,0}$ -- 6$_{-3,0}$ & 114.8 & 0.6 & 1.8 (0.4) & -59.0 (0.0) & 3.3 (0.0) & 527 (6) \\
 & 244337.983 & 9$_{1,1}$ -- 8$_{0,1}$ & 387.7 & 4.0 & 1.5 (0.3) & -59.4 (0.1) & 4.4 (0.1) & 329 (9) \\
 & 245094.503 & 18$_{-6,1}$ -- 17$_{-7,1}$ & 881.0 & 2.1 & 0.2 (0.0) & -58.6 (0.2) & 3.0 (0.4) & 67 (7) \\
 & 247161.950 & 16$_{2,0}$ -- 15$_{3,0}$ & 330.2 & 2.6 & 1.9 (0.4) & -59.3 (0.0) & 3.5 (0.1) & 518 (12) \\
 & 247840.050 & 12$_{-2,1}$ -- 13$_{-3,1}$ & 537.2 & 6.3 & 1.7 (0.3) & -58.9 (0.0) & 3.7 (0.1) & 434 (8) \\
 & 247968.119 & 23$_{1,0}$ -- 23$_{0,0}$ & 653.5 & 4.4 & 1.0 (0.2) & -59.0 (0.0) & 4.1 (0.1) & 228 (5) \\
 & 248854.996 & 15$_{-1,1}$ -- 16$_{-2,1}$ & 683.3 & 1.3 & 0.4 (0.1) & -58.7 (0.1) & 3.6 (0.2) & 107 (5) \\
 & 249004.019 & 10$_{7,1}$ -- 9$_{6,1}$ & 700.5 & 3.6 & 0.5 (0.1) & -58.7 (0.1) & 3.9 (0.2) & 114 (6) \\
 & 249192.836 & 16$_{-3,0}$ -- 15$_{-4,0}$ & 370.4 & 2.5 & 1.6 (0.3) & -59.0 (0.0) & 4.0 (0.1) & 376 (6) \\
 & 250970.042 & 17$_{3,1}$ -- 18$_{4,1}$ & 763.1 & 7.7 & 1.0 (0.2) & -59.0 (0.1) & 3.7 (0.1) & 250 (7) \\
 & 254015.377 & 2$_{0,0}$ -- 1$_{-1,0}$ & 12.2 & 1.9 & 3.1 (0.6) & -58.9 (0.3) & 4.1 (0.3) & 704 (104) \\
 & 254419.419 & 11$_{5,0}$ -- 12$_{4,0}$ & 281.3 & 1.8 & 1.8 (0.4) & -58.7 (0.0) & 3.2 (0.1) & 536 (9) \\
 & 259581.398 & 24$_{1,0}$ -- 24$_{0,0}$ & 709.1 & 4.9 & 0.7 (0.1) & -59.0 (0.1) & 3.4 (0.2) & 191 (8) \\
 & 261061.320 & 21$_{-4,0}$ -- 20$_{-5,0}$ & 615.7 & 3.0 & 1.0 (0.2) & -58.8 (0.1) & 2.9 (0.3) & 327 (29) \\
 & 261704.409 & 12$_{6,0}$ -- 13$_{5,0}$ & 351.9 & 1.8 & 1.1 (0.2) & -59.2 (0.1) & 3.4 (0.2) & 301 (19) \\
 & 261805.675 & 2$_{1,0}$ -- 1$_{0,0}$ & 20.1 & 5.6 & 5.3 (1.1) & -58.2 (0.0) & 4.4 (0.1) & 1132 (15) \\
 & 264732.426 & 13$_{2,1}$ -- 12$_{3,1}$ & 602.6 & 0.5 & 0.2 (0.1) & -58.6 (0.2) & 3.0 (0.4) & 75 (10) \\
 & 265289.562 & 6$_{1,0}$ -- 5$_{2,0}$ & 61.9 & 2.6 & 1.5 (0.3) & -59.0 (0.0) & 2.9 (0.2) & 497 (40) \\
 & 266838.148 & 5$_{2,0}$ -- 4$_{1,0}$ & 49.2 & 7.7 & 3.9 (0.8) & -57.9 (0.0) & 3.5 (0.1) & 1052 (24) \\
 & 267887.317 & 24$_{5,0}$ -- 23$_{6,0}$ & 823.6 & 3.2 & 0.5 (0.1) & -59.1 (0.2) & 3.5 (0.4) & 127 (13) \\
 & 268743.954 & 9$_{-5,0}$ -- 10$_{-4,0}$ & 220.5 & 1.8 & 1.8 (0.4) & -59.2 (0.1) & 3.8 (0.2) & 449 (23) \\
 & 271222.675 & 26$_{3,0}$ -- 25$_{4,0}$ & 854.6 & 3.8 & 0.7 (0.1) & -58.7 (0.2) & 4.3 (0.4) & 154 (11) \\
 & 271933.603 & 25$_{1,0}$ -- 25$_{0,0}$ & 767.0 & 5.5 & 0.9 (0.2) & -59.0 (0.1) & 3.6 (0.3) & 230 (18) \\
 & 274022.001 & 24$_{-7,0}$ -- 25$_{-6,0}$ & 947.6 & 3.1 & 0.2 (0.1) & -58.9 (0.2) & 3.8 (0.5) & 60 (6) \\
\hline
\end{tabular}
}
\label{taba1}
\end{table*}

\begin{table*}
\caption{Methanol}
\resizebox{1.75\columnwidth}{!}{
\begin{tabular}{lrcrrrrrr}
\hline\hline
Species & Frequency & Quantum numbers & $E_{up}$ & $A_{ul}$ & Flux & V$_{lsr}$ & $\Delta$V & $T_{peak}$ \\
 & (MHz) &  & (K) & (10$^{-5 }$\,s$^{-1}$) & (K\,km\,s$^{-1}$) & (km\,s$^{-1}$) & (km\,s$^{-1}$) & (mK) \\
 \hline
CH$_3$OH-A & 215302.206 & 6$_{1,+,1}$ -- 7$_{2,+,1}$ & 373.8 & 4.2 & 2.4 (0.5) & -59.1 (0.0) & 3.6 (0.1) & 637 (8) \\
 & 217299.205 & 6$_{1,-,1}$ -- 7$_{2,-,1}$ & 373.9 & 4.3 & 1.8 (0.4) & -59.0 (0.0) & 3.1 (0.1) & 530 (23) \\
 & 231281.110 & 10$_{2,-,0}$ -- 9$_{3,-,0}$ & 165.3 & 1.8 & 1.7 (0.3) & -59.0 (0.0) & 3.5 (0.1) & 460 (18) \\
 & 232418.521 & 10$_{2,+,0}$ -- 9$_{3,+,0}$ & 165.4 & 1.9 & 1.6 (0.3) & -58.9 (0.0) & 3.0 (0.1) & 518 (16) \\
 & 232783.446 & 18$_{3,+,0}$ -- 17$_{4,+,0}$ & 446.5 & 2.2 & 2.4 (0.5) & -59.2 (0.0) & 3.8 (0.1) & 589 (7) \\
 & 233795.666 & 18$_{3,-,0}$ -- 17$_{4,-,0}$ & 446.6 & 2.2 & 2.4 (0.5) & -59.2 (0.0) & 3.6 (0.0) & 617 (7) \\
 & 234683.370 & 4$_{2,-,0}$ -- 5$_{1,-,0}$ & 60.9 & 1.8 & 3.7 (0.7) & -58.9 (0.0) & 3.6 (0.0) & 955 (9) \\
 & 243915.788 & 5$_{1,-,0}$ -- 4$_{1,-,0}$ & 49.7 & 6.0 & 6.1 (1.2) & -58.0 (0.0) & 4.2 (0.0) & 1345 (9) \\
 & 246873.301 & 19$_{3,-,0}$ -- 19$_{2,+,0}$ & 490.7 & 8.0 & 1.5 (0.3) & -59.2 (0.0) & 3.3 (0.1) & 442 (9) \\
 & 247228.587 & 4$_{2,+,0}$ -- 5$_{1,+,0}$ & 60.9 & 2.2 & 3.5 (0.7) & -59.0 (0.0) & 3.7 (0.1) & 876 (14) \\
 & 247610.918 & 18$_{3,-,0}$ -- 18$_{2,+,0}$ & 446.6 & 8.1 & 1.5 (0.3) & -59.2 (0.0) & 3.3 (0.1) & 437 (9) \\
 & 248282.424 & 17$_{3,-,0}$ -- 17$_{2,+,0}$ & 404.8 & 8.1 & 1.7 (0.3) & -59.2 (0.0) & 3.6 (0.1) & 444 (6) \\
 & 248885.468 & 16$_{3,-,0}$ -- 16$_{2,+,0}$ & 365.4 & 8.2 & 1.8 (0.4) & -59.2 (0.0) & 3.4 (0.1) & 482 (9) \\
 & 249419.924 & 15$_{3,-,0}$ -- 15$_{2,+,0}$ & 328.3 & 8.2 & 1.8 (0.4) & -59.2 (0.0) & 3.5 (0.1) & 479 (11) \\
 & 249443.301 & 7$_{4,-,0}$ -- 8$_{3,-,0}$ & 145.3 & 1.5 & 1.6 (0.3) & -59.1 (0.0) & 4.0 (0.1) & 375 (8) \\
 & 249451.842 & 7$_{4,+,0}$ -- 8$_{3,+,0}$ & 145.3 & 1.5 & 1.5 (0.3) & -59.1 (0.0) & 3.9 (0.1) & 372 (8) \\
 & 249887.467 & 14$_{3,-,0}$ -- 14$_{2,+,0}$ & 293.5 & 8.2 & 1.5 (0.3) & -59.2 (0.0) & 3.1 (0.1) & 443 (14) \\
 & 250291.181 & 13$_{3,-,0}$ -- 13$_{2,+,0}$ & 261.0 & 8.2 & 2.0 (0.4) & -59.1 (0.0) & 3.6 (0.1) & 526 (8) \\
 & 250506.853 & 11$_{0,+,0}$ -- 10$_{1,+,0}$ & 153.1 & 4.2 & 2.6 (0.5) & -58.6 (0.0) & 4.5 (0.1) & 535 (7) \\
 & 250635.200 & 12$_{3,-,0}$ -- 12$_{2,+,0}$ & 230.8 & 8.2 & 2.3 (0.5) & -59.3 (0.0) & 3.7 (0.1) & 590 (8) \\
 & 250924.398 & 11$_{3,-,0}$ -- 11$_{2,+,0}$ & 203.0 & 8.2 & 2.5 (0.5) & -59.2 (0.0) & 3.5 (0.0) & 673 (5) \\
 & 251164.108 & 10$_{3,-,0}$ -- 10$_{2,+,0}$ & 177.5 & 8.2 & 2.7 (0.5) & -59.1 (0.0) & 3.5 (0.1) & 711 (12) \\
 & 251517.309 & 8$_{3,-,0}$ -- 8$_{2,+,0}$ & 133.4 & 7.9 & 3.1 (0.6) & -59.0 (0.0) & 3.5 (0.1) & 835 (10) \\
 & 251641.787 & 7$_{3,-,0}$ -- 7$_{2,+,0}$ & 114.8 & 7.7 & 3.2 (0.6) & -59.0 (0.0) & 3.8 (0.1) & 797 (9) \\
 & 251738.437 & 6$_{3,-,0}$ -- 6$_{2,+,0}$ & 98.5 & 7.4 & 3.4 (0.7) & -58.8 (0.0) & 4.0 (0.0) & 800 (8) \\
 & 251811.956 & 5$_{3,-,0}$ -- 5$_{2,+,0}$ & 84.6 & 7.0 & 3.6 (0.7) & -58.8 (0.0) & 4.3 (0.1) & 790 (17) \\
 & 251866.524 & 4$_{3,-,0}$ -- 4$_{2,+,0}$ & 73.0 & 6.1 & 3.1 (0.6) & -58.6 (0.0) & 4.1 (0.1) & 717 (8) \\
 & 251890.886 & 5$_{3,+,0}$ -- 5$_{2,-,0}$ & 84.6 & 7.0 & 3.3 (0.7) & -58.5 (0.0) & 3.8 (0.1) & 818 (8) \\
 & 251895.728 & 6$_{3,+,0}$ -- 6$_{2,-,0}$ & 98.5 & 7.5 & 3.4 (0.7) & -58.6 (0.0) & 3.8 (0.1) & 817 (7) \\
 & 251900.452 & 4$_{3,+,0}$ -- 4$_{2,-,0}$ & 73.0 & 6.1 & 3.4 (0.7) & -58.5 (0.0) & 4.0 (0.1) & 781 (9) \\
 & 251905.729 & 3$_{3,-,0}$ -- 3$_{2,+,0}$ & 63.7 & 4.4 & 3.4 (0.7) & -58.7 (0.0) & 4.2 (0.1) & 629 (8) \\
 & 251917.065 & 3$_{3,+,0}$ -- 3$_{2,-,0}$ & 63.7 & 4.4 & 3.0 (0.6) & -58.8 (0.0) & 4.1 (0.1) & 683 (9) \\
 & 251923.701 & 7$_{3,+,0}$ -- 7$_{2,-,0}$ & 114.8 & 7.8 & 3.2 (0.7) & -58.8 (0.0) & 3.8 (0.0) & 793 (6) \\
 & 251984.837 & 8$_{3,+,0}$ -- 8$_{2,-,0}$ & 133.4 & 8.0 & 3.1 (0.6) & -59.0 (0.0) & 3.7 (0.1) & 780 (10) \\
 & 252090.409 & 9$_{3,+,0}$ -- 9$_{2,-,0}$ & 154.2 & 8.1 & 3.2 (0.6) & -59.0 (0.0) & 3.7 (0.1) & 799 (9) \\
 & 252252.849 & 10$_{3,+,0}$ -- 10$_{2,-,0}$ & 177.5 & 8.3 & 2.9 (0.6) & -59.1 (0.0) & 3.6 (0.0) & 749 (8) \\
 & 252485.675 & 11$_{3,+,0}$ -- 11$_{2,-,0}$ & 203.0 & 8.4 & 2.7 (0.5) & -59.1 (0.0) & 3.6 (0.0) & 708 (8) \\
 & 252803.388 & 12$_{3,+,0}$ -- 12$_{2,-,0}$ & 230.8 & 8.4 & 2.5 (0.5) & -59.2 (0.0) & 3.5 (0.1) & 677 (9) \\
 & 253221.376 & 13$_{3,+,0}$ -- 13$_{2,-,0}$ & 261.0 & 8.5 & 2.4 (0.5) & -59.2 (0.0) & 3.4 (0.1) & 660 (9) \\
 & 253755.809 & 14$_{3,+,0}$ -- 14$_{2,-,0}$ & 293.5 & 8.6 & 2.5 (0.5) & -59.2 (0.0) & 3.4 (0.1) & 694 (11) \\
 & 254423.520 & 15$_{3,+,0}$ -- 15$_{2,-,0}$ & 328.3 & 8.7 & 2.6 (0.5) & -59.2 (0.0) & 3.7 (0.1) & 674 (10) \\
 & 255241.888 & 16$_{3,+,0}$ -- 16$_{2,-,0}$ & 365.4 & 8.8 & 2.2 (0.4) & -59.2 (0.0) & 3.3 (0.1) & 615 (8) \\
 & 256228.714 & 17$_{3,+,0}$ -- 17$_{2,-,0}$ & 404.8 & 9.0 & 1.9 (0.4) & -59.3 (0.0) & 3.3 (0.1) & 545 (10) \\
 & 257402.086 & 18$_{3,+,0}$ -- 18$_{2,-,0}$ & 446.5 & 9.1 & 2.8 (0.6) & -59.9 (0.0) & 4.3 (0.1) & 613 (14) \\
 & 258780.248 & 19$_{3,+,0}$ -- 19$_{2,-,0}$ & 490.6 & 9.3 & 2.1 (0.4) & -59.4 (0.0) & 3.9 (0.1) & 502 (11) \\
 & 263793.875 & 5$_{1,+,1}$ -- 6$_{2,+,1}$ & 360.0 & 8.2 & 2.7 (0.5) & -58.5 (0.1) & 4.5 (0.1) & 566 (13) \\
 & 265224.426 & 5$_{1,-,1}$ -- 6$_{2,-,1}$ & 360.0 & 8.3 & 2.2 (0.4) & -59.0 (0.0) & 3.6 (0.1) & 569 (9) \\
 & 267406.071 & 17$_{1,+,0}$ -- 16$_{2,+,0}$ & 366.3 & 4.2 & 5.3 (1.1) & -56.9 (0.1) & 7.3 (0.2) & 681 (9) \\
 & 260381.463 & 20$_{3,+,0}$ -- 20$_{2,-,0}$ & 536.9 & 9.4 & 2.3 (0.5) & -59.8 (0.1) & 4.8 (0.1) & 438 (11) \\
 & 246074.605 & 20$_{3,-,0}$ -- 20$_{2,+,0}$ & 537.0 & 8.0 & 1.7 (0.3) & -59.5 (0.0) & 3.7 (0.1) & 417 (10) \\
 & 229939.095 & 19$_{5,-,0}$ -- 20$_{4,-,0}$ & 578.6 & 2.1 & 1.3 (0.3) & -59.0 (0.1) & 3.9 (0.2) & 315 (12) \\
 & 229864.121 & 19$_{5,+,0}$ -- 20$_{4,+,0}$ & 578.6 & 2.1 & 1.4 (0.3) & -58.8 (0.1) & 4.8 (0.2) & 281 (9) \\
 & 262223.872 & 21$_{3,+,0}$ -- 21$_{2,-,0}$ & 585.6 & 9.7 & 1.3 (0.3) & -59.3 (0.0) & 3.5 (0.1) & 350 (10) \\
 & 245223.019 & 21$_{3,-,0}$ -- 21$_{2,+,0}$ & 585.7 & 7.9 & 1.9 (0.4) & -58.6 (0.2) & 4.3 (0.2) & 408 (13) \\
 & 264325.354 & 22$_{3,+,0}$ -- 22$_{2,-,0}$ & 636.6 & 9.9 & 1.5 (0.3) & -59.3 (0.1) & 4.3 (0.2) & 325 (14) \\
 & 244330.372 & 22$_{3,-,0}$ -- 22$_{2,+,0}$ & 636.8 & 7.8 & 1.5 (0.3) & -59.3 (0.0) & 3.9 (0.1) & 362 (8) \\
 & 271562.485 & 17$_{2,+,1}$ -- 16$_{1,+,1}$ & 652.6 & 6.5 & 1.4 (0.3) & -59.2 (0.1) & 4.4 (0.2) & 307 (12) \\
 & 259273.686 & 17$_{2,-,1}$ -- 16$_{1,-,1}$ & 652.7 & 5.6 & 1.1 (0.2) & -59.0 (0.0) & 3.8 (0.1) & 270 (6) \\
 & 266703.383 & 23$_{3,+,0}$ -- 23$_{2,-,0}$ & 689.9 & 10.2 & 1.0 (0.2) & -59.3 (0.0) & 3.2 (0.1) & 306 (7) \\
 & 266872.190 & 14$_{6,+,1}$ -- 15$_{5,+,1}$ & 711.0 & 3.1 & 0.9 (0.2) & -59.0 (0.1) & 3.6 (0.2) & 234 (10) \\
 & 269374.884 & 24$_{3,+,0}$ -- 24$_{2,-,0}$ & 745.5 & 10.5 & 1.2 (0.3) & -59.2 (0.1) & 3.7 (0.3) & 310 (21) \\
 & 217642.677 & 15$_{6,-,1}$ -- 16$_{5,-,1}$ & 745.6 & 1.9 & 1.1 (0.2) & -58.8 (0.1) & 4.4 (0.1) & 234 (7) \\
 & 272356.098 & 25$_{3,+,0}$ -- 25$_{2,-,0}$ & 803.4 & 10.9 & 1.2 (0.2) & -59.2 (0.0) & 3.6 (0.2) & 308 (8) \\
 & 233916.950 & 13$_{3,-,2}$ -- 14$_{4,-,2}$ & 868.5 & 1.4 & 0.4 (0.1) & -58.7 (0.1) & 3.6 (0.2) & 115 (4) \\
 & 233917.018 & 13$_{3,+,2}$ -- 14$_{4,+,2}$ & 868.5 & 1.4 & 0.4 (0.1) & -58.7 (0.1) & 3.5 (0.3) & 111 (7) \\
 & 268345.816 & 20$_{4,-,1}$ -- 21$_{5,-,1}$ & 967.6 & 5.3 & 0.8 (0.2) & -58.9 (0.1) & 4.1 (0.3) & 189 (14) \\
 & 268346.226 & 20$_{4,+,1}$ -- 21$_{5,+,1}$ & 967.6 & 5.3 & 0.8 (0.2) & -58.5 (0.2) & 4.1 (0.4) & 189 (14) \\
\hline
\end{tabular}
}
\label{taba2}
\end{table*}

\begin{table*}
\caption{\label{13CH3OH-lines}$^{13}$C Methanol}
\begin{tabular}{lrrrrrrrr}

\hline\hline
Species & Frequency & Quantum numbers & $E_{up}$ & $A_{ul}$ & Flux & V$_{lsr}$ & $\Delta$V & $T_{peak}$ \\
 & (MHz) &  & (K) & (10$^{-5 }$\,s$^{-1}$) & (K\,km\,s$^{-1}$) & (km\,s$^{-1}$) & (km\,s$^{-1}$) & (mK) \\
\hline
$^{13}$CH$_3$OH & 212775.66 & 1$_{1,0,0}$ -- 0$_{0,0,0}$ & 23.3 & 3.3 & 296 (60) & -58.8 (0.1) & 2.7 (0.2) & 104 (8) \\
 & 215707.92 & 8$_{4,5,-,0}$ -- 9$_{3,6,-,0}$ & 162.3 & 1.1 & 134 (33) & -58.7 (0.3) & 2.5 (0.7) & 51 (12) \\
 & 215722.48 & 8$_{4,4,+,0}$ -- 9$_{3,7,+,0}$ & 162.3 & 1.1 & 222 (45) & -58.9 (0.1) & 3.3 (0.2) & 63 (2) \\
 & 216370.39 & 10$_{2,9,-,0}$ -- 9$_{3,6,-,0}$ & 162.4 & 1.5 & 540 (113) & -59.5 (0.2) & 4.9 (0.5) & 103 (8) \\
 & 234011.58 & 5$_{1,5,+,0}$ -- 4$_{1,4,+,0}$ & 48.3 & 5.3 & 1093 (220) & -58.8 (0.0) & 3.5 (0.2) & 290 (12) \\
 & 235881.17 & 5$_{0,5,0}$ -- 4$_{0,4,0}$ & 47.1 & 5.6 & 1044 (209) & -58.8 (0.0) & 3.2 (0.1) & 307 (7) \\
 & 235938.22 & 5$_{-1,5,0}$ -- 4$_{-1,4,0}$ & 39.6 & 5.4 & 1092 (221) & -58.8 (0.1) & 3.2 (0.2) & 318 (18) \\
 & 235960.37 & 5$_{0,5,+,0}$ -- 4$_{0,4,+,0}$ & 34.0 & 5.6 & 1032 (207) & -58.7 (0.0) & 3.1 (0.0) & 316 (4) \\
 & 235971.07 & 5$_{4,2,-,0}$ -- 4$_{4,1,-,0}$ & 114.8 & 2.0 & 498 (103) & -58.6 (0.1) & 2.7 (0.3) & 176 (19) \\
 & 235971.07 & 5$_{4,1,+,0}$ -- 4$_{4,0,+,0}$ & 114.8 & 2.0 & 502 (101) & -58.6 (0.0) & 2.7 (0.1) & 175 (6) \\
 & 235978.62 & 5$_{-4,2,0}$ -- 4$_{-4,1,0}$ & 122.3 & 2.0 & 307 (64) & -58.9 (0.2) & 2.6 (0.6) & 110 (19) \\
 & 235997.23 & 5$_{3,3,+,0}$ -- 4$_{3,2,+,0}$ & 84.0 & 3.6 & 1772 (360) & -57.9 (0.1) & 4.6 (0.3) & 360 (17) \\
 & 235997.23 & 5$_{3,2,-,0}$ -- 4$_{3,1,-,0}$ & 84.0 & 3.6 & 1771 (360) & -57.9 (0.1) & 4.6 (0.3) & 360 (18) \\
 & 236008.39 & 5$_{2,4,-,0}$ -- 4$_{2,3,-,0}$ & 71.8 & 4.8 & 1390 (287) & -57.6 (0.2) & 5.1 (0.5) & 258 (19) \\
 & 236006.10 & 5$_{3,2,0}$ -- 4$_{3,1,0}$ & 81.9 & 3.6 & 1459 (299) & -60.6 (0.2) & 5.4 (0.5) & 253 (15) \\
 & 236016.55 & 5$_{-3,3,0}$ -- 4$_{-3,2,0}$ & 96.9 & 3.6 & 428 (90) & -58.7 (0.1) & 2.5 (0.4) & 162 (20) \\
 & 236041.40 & 5$_{1,4,0}$ -- 4$_{1,3,0}$ & 55.0 & 5.5 & 1002 (202) & -58.7 (0.0) & 3.5 (0.1) & 267 (8) \\
 & 236049.52 & 5$_{2,3,+,0}$ -- 4$_{2,2,+,0}$ & 71.8 & 4.8 & 1300 (260) & -58.5 (0.0) & 3.7 (0.1) & 332 (5) \\
 & 236062.00 & 5$_{-2,4,0}$ -- 4$_{-2,3,0}$ & 60.0 & 4.7 & 1971 (397) & -59.3 (0.1) & 4.3 (0.1) & 432 (13) \\
 & 236062.85 & 5$_{2,3,0}$ -- 4$_{2,2,0}$ & 56.3 & 4.7 & 1973 (397) & -58.2 (0.1) & 4.3 (0.2) & 431 (18) \\
 & 248654.97 & 10$_{-3,8,0}$ -- 11$_{-2,10,0}$ & 187.5 & 2.6 & 234 (48) & -58.4 (0.1) & 2.4 (0.2) & 91 (7) \\
 & 250125.69 & 5$_{3,2,0}$ -- 6$_{2,4,0}$ & 81.9 & 1.6 & 164 (34) & -58.5 (0.1) & 2.8 (0.3) & 54 (5) \\
 & 250784.61 & 2$_{0,2,0}$ -- 1$_{-1,1,0}$ & 19.9 & 1.8 & 179 (37) & -58.7 (0.1) & 2.8 (0.3) & 61 (6) \\
 & 254321.72 & 4$_{2,2,+,0}$ -- 5$_{1,5,+,0}$ & 60.5 & 2.3 & 365 (74) & -58.6 (0.1) & 3.1 (0.2) & 112 (6) \\
 & 254509.36 & 10$_{3,7,-,0}$ -- 10$_{2,8,+,0}$ & 174.6 & 8.5 & 1190 (239) & -58.7 (0.0) & 3.3 (0.1) & 340 (10) \\
 & 254959.40 & 7$_{3,4,-,0}$ -- 7$_{2,5,+,0}$ & 113.5 & 8.1 & 1042 (209) & -58.8 (0.0) & 3.1 (0.1) & 314 (9) \\
 & 255050.97 & 6$_{3,3,-,0}$ -- 6$_{2,4,+,0}$ & 97.6 & 7.8 & 1569 (315) & -58.6 (0.1) & 3.9 (0.1) & 383 (12) \\
 & 251796.08 & 17$_{3,14,-,0}$ -- 17$_{2,15,+,0}$ & 396.5 & 8.6 & 465 (94) & -58.6 (0.1) & 3.3 (0.1) & 131 (5) \\
 & 259036.49 & 17$_{3,15,+,0}$ -- 17$_{2,16,-,0}$ & 396.5 & 9.1 & 667 (135) & -57.8 (0.1) & 4.1 (0.2) & 152 (6) \\
 & 252870.23 & 15$_{3,12,-,0}$ -- 15$_{2,13,+,0}$ & 321.8 & 8.7 & 697 (140) & -58.7 (0.0) & 3.4 (0.1) & 191 (3) \\
 & 257421.79 & 15$_{3,13,+,0}$ -- 15$_{2,14,-,0}$ & 321.8 & 9.0 & 607 (122) & -58.5 (0.0) & 3.3 (0.1) & 175 (5) \\
 & 256826.57 & 14$_{3,12,+,0}$ -- 14$_{2,13,-,0}$ & 287.8 & 8.9 & 758 (153) & -58.7 (0.1) & 3.2 (0.2) & 221 (9) \\
 & 273962.69 & 11$_{5,7,0}$ -- 12$_{4,8,0}$ & 286.2 & 2.2 & 103 (24) & -58.7 (0.2) & 1.8 (0.5) & 52 (14) \\
 & 269530.49 & 15$_{-1,15,0}$ -- 14$_{-2,13,0}$ & 277.0 & 3.1 & 324 (81) & -58.0 (0.3) & 2.8 (0.8) & 107 (25) \\
\hline
\end{tabular}
\label{taba3}
\end{table*}

\begin{table*}
\caption{\label{deuterated-lines}Deuterated Methanol}
\begin{tabular}{lrrrrrrrr}

\hline\hline
Species & Frequency & Quantum numbers & $E_{up}$ & $A_{ul}$ & Flux & V$_{lsr}$ & $\Delta$V & $T_{peak}$ \\
 & (MHz) &  & (K) & (10$^{-5 }$\,s$^{-1}$) & (K\,km\,s$^{-1}$) & (km\,s$^{-1}$) & (km\,s$^{-1}$) & (mK) \\
\hline
CH$_2$DOH & 234471.03 & 8$_{2,6,0}$ -- 8$_{1,7,0}$ & 93.7 & 8.4 & 171 (38) & -57.5 (0.2) & 2.8 (0.4) & 57 (6) \\
 & 255647.82 & 3$_{2,2,0}$ -- 3$_{1,3,0}$ & 29.0 & 6.3 & 101 (28) & -57.9 (0.3) & 2.3 (0.8) & 41 (11) \\
 & 256731.55 & 4$_{1,4,0}$ -- 3$_{0,3,0}$ & 25.2 & 6.9 & 126 (31) & -58.2 (0.3) & 3.4 (0.7) & 35 (6) \\
 & 258337.11 & 4$_{2,3,0}$ -- 4$_{1,4,0}$ & 37.6 & 7.0 & 101 (29) & -58.8 (0.3) & 2.9 (0.7) & 33 (7) \\
 & 264017.72 & 6$_{1,6,0}$ -- 5$_{1,5,0}$ & 48.4 & 5.8 & 182 (51) & -58.2 (0.3) & 2.5 (0.6) & 68 (14) \\
 & 265509.20 & 6$_{1,6,2}$ -- 5$_{1,5,2}$ & 67.5 & 7.5 & 126 (30) & -58.0 (0.2) & 2.5 (0.5) & 48 (7) \\
 & 265682.51 & 6$_{2,5,0}$ -- 6$_{1,6,0}$ & 61.2 & 8.2 & 126 (32) & -58.1 (0.2) & 3.0 (0.6) & 40 (7) \\
 & 267634.61 & 6$_{0,6,0}$ -- 5$_{0,5,0}$ & 45.0 & 6.3 & 187 (40) & -58.2 (0.1) & 2.3 (0.2) & 77 (6) \\
 & 267731.74 & 6$_{2,5,1}$ -- 5$_{2,4,1}$ & 71.6 & 7.1 & 163 (38) & -57.9 (0.1) & 2.4 (0.4) & 64 (8) \\
 & 267741.09 & 6$_{3,4,1}$ -- 5$_{3,3,1}$ & 89.9 & 5.9 & 258 (54) & -58.6 (0.1) & 3.2 (0.3) & 76 (5) \\
 & 267742.05 & 6$_{3,3,1}$ -- 5$_{3,2,1}$ & 89.9 & 5.9 & 261 (55) & -57.5 (0.1) & 3.3 (0.3) & 76 (5) \\
 &  270299.93  &   7$_{2,6,0}$ -- 7$_{1,70}$   &   76.2    &  8.8   &  117(31)   &  -57.8 (0.1)  &  2.6 (0.3)  &   43 (4) \\
 & 270734.57 & 6$_{1,5,1}$ -- 5$_{1,4,1}$ & 62.0 & 7.8 & 187 (41) & -58.3 (0.1) & 2.5 (0.3) & 69 (7) \\
\hline%
CH$_3$OD-E & 217132.73 & 9$_{1,8,0}$ -- 8$_{2,6,0}$ & 107.32 & 1.7 & 115 (28) & -58.0 (0.3) & 4.2 (0.9) & 26 (5) \\
 & 245143.16 & 5$_{-1,5,0}$ -- 4$_{0,4,0}$ & 37.27 & 5.9 & 212 (54) & -58.6 (0.3) & 3.8 (0.7) & 53 (12) \\
 & 252624.31 & 6$_{0,6,0}$ -- 5$_{-1,5,0}$ & 49.39 & 5.0 & 207 (47) & -57.6 (0.1) & 2.8 (0.3) & 70 (10) \\
 & 264576.81 & 10$_{1,9,0}$ -- 9$_{2,7,0}$ & 129.23 & 3.0 & 115 (31) & -59.0 (0.3) & 3.4 (1.0) & 31 (8) \\
 & 269844.66 & 4$_{-2,3,0}$ -- 4$_{-1,4,0}$ & 39.36 & 11.1 & 156 (46) & -58.9 (0.3) & 1.9 (0.4) & 75 (18) \\
 & 270581.30 & 5$_{-2,4,0}$ -- 5$_{-1,5,0}$ & 50.25 & 11.6 & 310 (65) & -58.6 (0.1) & 3.2 (0.3) & 90 (8) \\
 & 271079.19 & 6$_{-1,6,0}$ -- 5$_{-1,5,0}$ & 50.27 & 7.5 & 416 (88) & -58.4 (0.1) & 3.6 (0.3) & 109 (12) \\
 & 271417.31 & 6$_{0,6,0}$ -- 5$_{0,5,0}$ & 49.39 & 6.1 & 332 (73) & -57.9 (0.3) & 3.9 (0.8) & 80 (12) \\
 & 272328.66 & 6$_{2,4,0}$ -- 5$_{2,3,0}$ & 64.21 & 6.9 & 187 (48) & -58.3 (0.2) & 3.0 (0.5) & 58 (13) \\
 & 272922.56 & 6$_{1,5,0}$ -- 5$_{1,4,0}$ & 54.83 & 7.7 & 435 (101) & -57.5 (0.3) & 3.8 (0.6) & 108 (20) \\
\hline%
CH$_3$OD-A& 218156.16 & 14$_{1,13,0}$ -- 14$_{0,14,0}$ & 237.96 & 9.1 & 259 (54) & -57.5 (0.4) & 3.1 (0.7) & 79 (6) \\
 & 230105.09 & 5$_{1,4,0}$ -- 4$_{1,3,0}$ & 39.53 & 4.4 & 286 (77) & -58.2 (0.3) & 2.8 (0.6) & 94 (21) \\
 & 232077.53 & 15$_{1,14,0}$ -- 15$_{0,15,0}$ & 270.96 & 10.5 & 740 (150) & -56.9 (0.2) & 3.9 (0.4) & 180 (8) \\
 & 246522.69 & 8$_{0,8,0}$ -- 7$_{1,7,0}$ & 78.24 & 8.6 & 596 (127) & -58.9 (0.7) & 6.3 (2.2) & 89 (13) \\
 & 265235.75 & 3$_{1,3,0}$ -- 2$_{0,2,0}$ & 19.26 & 13.2 & 310 (66) & -58.4 (0.1) & 3.1 (0.3) & 94 (10) \\
 & 271704.94 & 6$_{0,6,0}$ -- 5$_{0,5,0}$ & 45.67 & 7.6 & 244 (72) & -58.8 (0.2) & 2.8 (0.4) & 81 (24) \\
 & 271843.69 & 14$_{2,12,0}$ -- 14$_{1,13,0}$ & 251 & 16.5 & 247 (62) & -58.5 (0.4) & 3.4 (0.8) & 69 (15) \\
 & 272004.50 & 6$_{2,5,0}$ -- 5$_{2,4,0}$ & 67.49 & 6.9 & 295 (67) & -58.1 (0.3) & 3.3 (0.7) & 83 (13) \\
 & 272138.36 & 6$_{3,3,0}$ -- 5$_{3,2,0}$ & 86.73 & 6.0 & 254 (60) & -57.8 (0.3) & 3.9 (0.9) & 61 (13) \\
 & 272417.25 & 6$_{2,4,0}$ -- 5$_{2,3,0}$ & 67.53 & 7.0 & 323 (72) & -58.0 (0.9) & 4.6 (0.9) & 66 (11) \\
\hline\hline%
\end{tabular}
\label{taba4}
\end{table*}

\section{Partition function of CH$_3$OD}
\label{appB}
Table~\ref{taba5} gives the respective CH$_3$OD-A and CH$_3$OD-E rotation partition functions we use in the present study.

\begin{table}[!htp]
\caption{Rotational partition function for CH$_3$OD}
\label{taba5}
\begin{center}
\begin{tabular}{ll|ll}
\hline\hline
\multicolumn{2}{c}{CH$_3$OD-A} &   \multicolumn{2}{c}{CH$_3$OD-E} \\
\hline
T [K] & Q & T [K] & Q\\
\hline
  9.375&18.1&9.375&15.5\\
 18.75 &52.3&18.75 &50.8\\
 37.5&153.5&37.5&153.2\\
 75&458.6&75&458.6\\
150&1537.5&150&1537.1 \\
225&3319.7&225&3316.7 \\
300 &5739.8&300&5727.9 \\
\hline\hline
\end{tabular}
\end{center}
\end{table}

\section{IRAM-30~m observations overlaid with our LTE modeling for $^{13}$CH$_3$OH, CH$_3$OD-A and CH$_3$OD-E. }
\label{appC}
\begin{figure*}[!htp]
\center
\includegraphics[width=2\columnwidth]{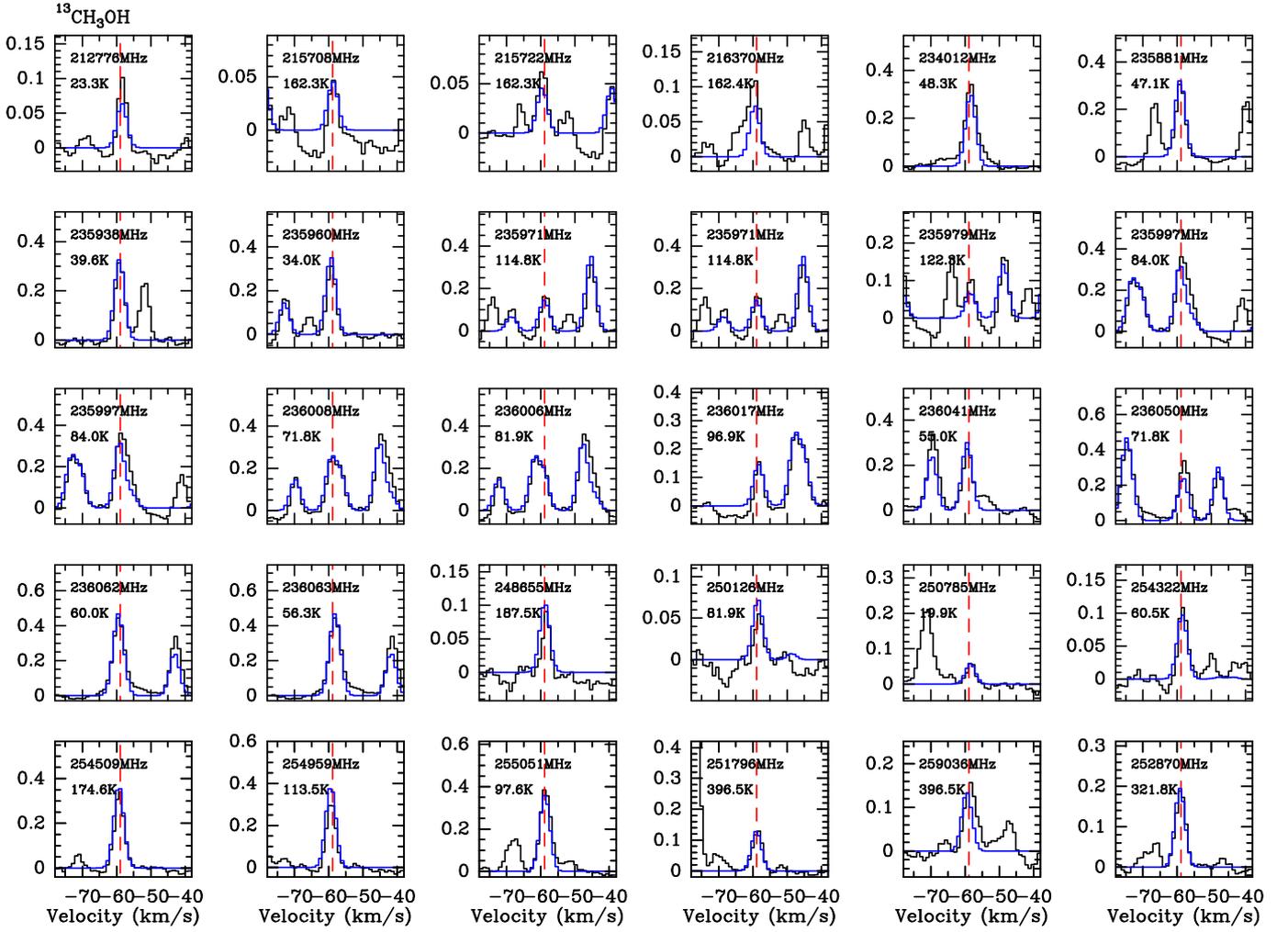}
\caption{\label{fig:c1}{Montage of detected transitions associated with the methanol $^{13}$C isotopologue  towards NGC~7538-IRS1. Our LTE modelling is displayed in blue. Intensities are expressed in unit of $T_{mb}$. The red dashed line marks the ambient cloud velocity $v_{LSR} = -58.9\kms$.}}
\end{figure*}

\begin{figure*}[!htp]
\center
\includegraphics[width=2\columnwidth]{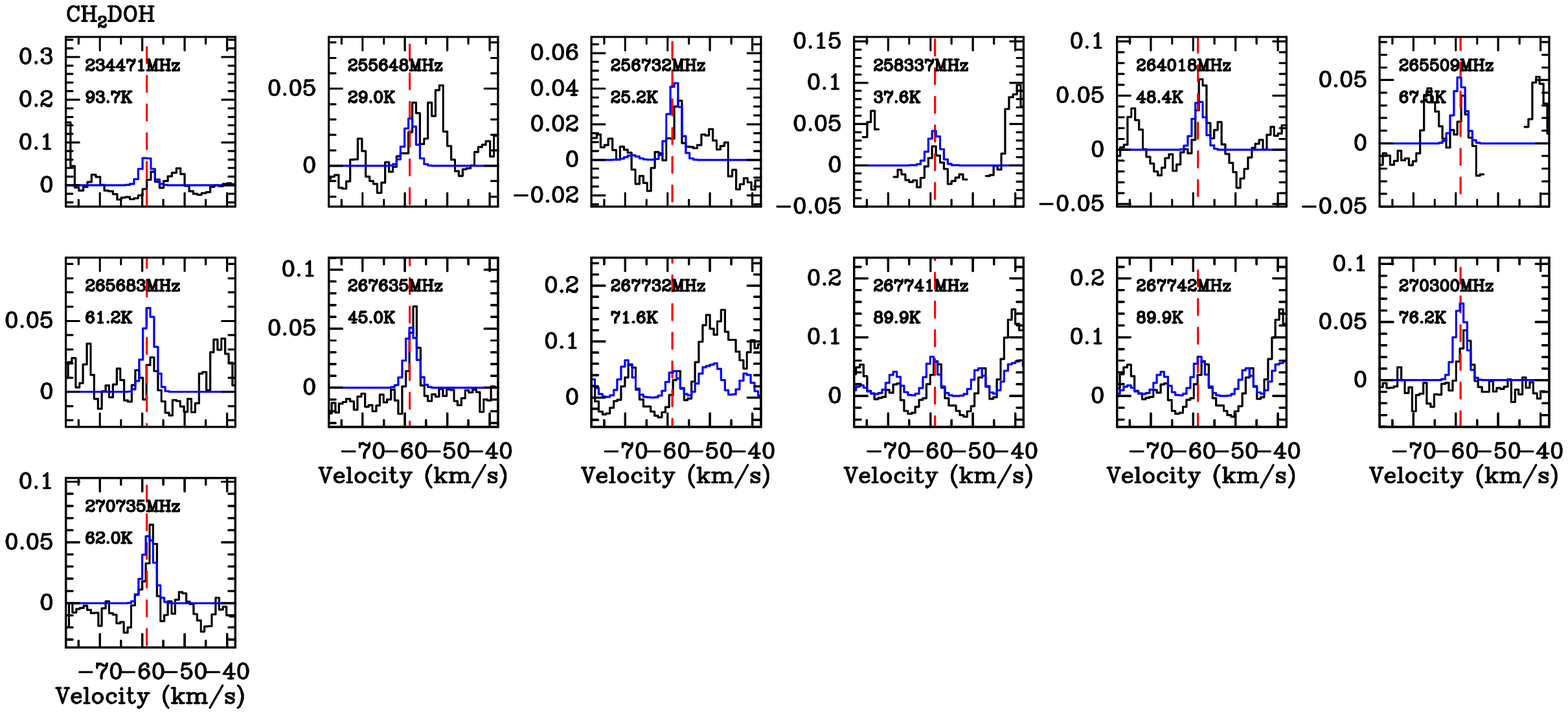}
\caption{\label{fig:c2}{Montage of detected transitions associated with the deuterated CH$_2$DOH methanol  towards NGC~7538-IRS1. Our LTE modelling is displayed in blue. Intensities are expressed in unit of $T_{mb}$. The red dashed line marks the ambient cloud velocity $v_{LSR} = -58.9\kms$.}}
\end{figure*}

\begin{figure*}[!htp]
\center
\includegraphics[width=2\columnwidth]{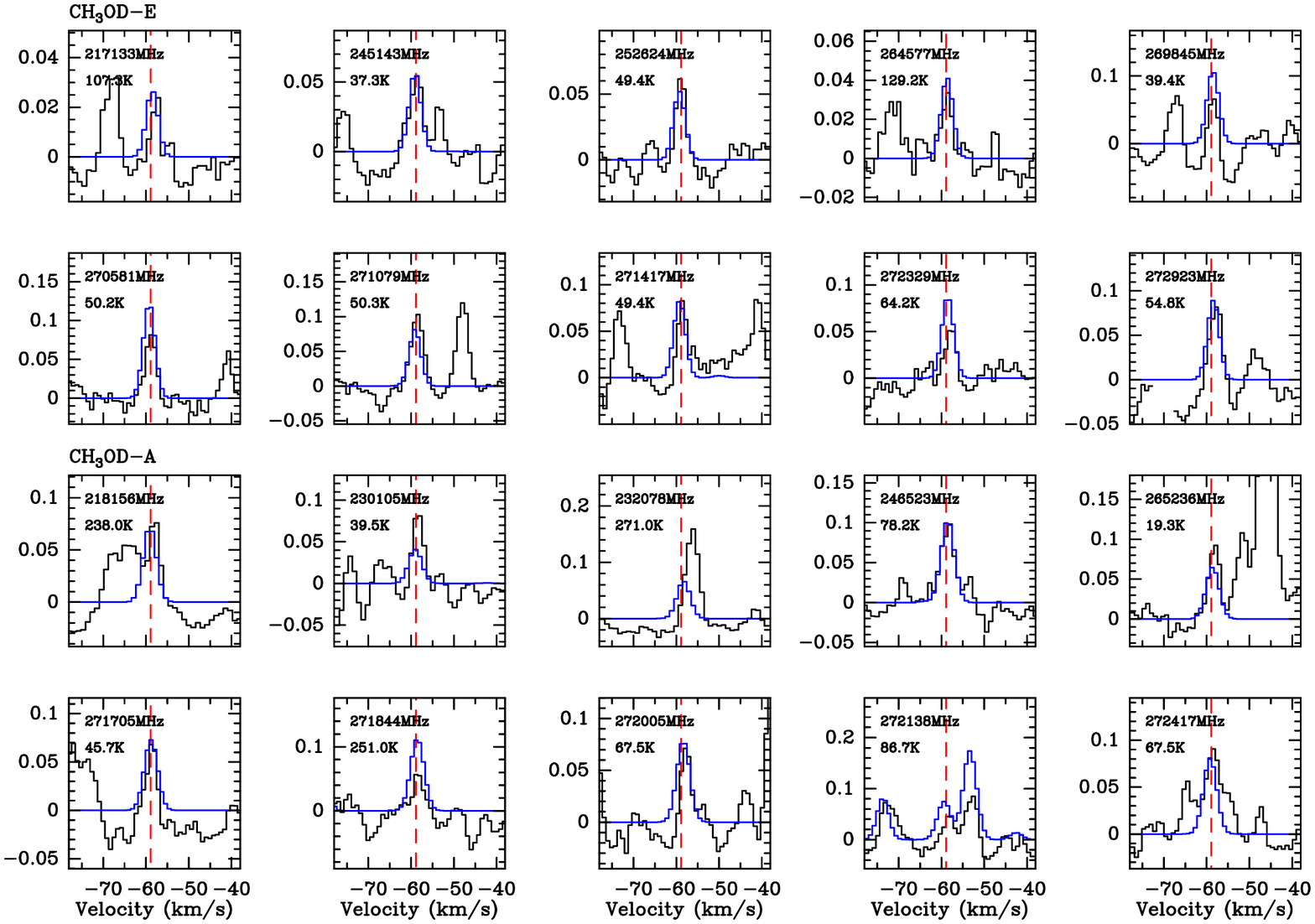}
\caption{\label{fig:c3}{Montage of detected transitions associated with the CH$_3$OD, -A and -E forms towards NGC~7538-IRS1. Our LTE modelling is displayed in blue. Intensities are expressed in unit of $T_{mb}$. The red dashed line marks the ambient cloud velocity $v_{LSR} = -58.9\kms$.}}
\end{figure*}

{Figures~\ref{fig:c1} to \ref{fig:c3} show a montage of the detected $^{13}$CH$_3$OH, CH$_2$DOH, CH$_3$OD-A and CH$_3$OD-E  transitions (see Tables \ref{taba3} to \ref{taba4} for the observational line parameters) along with our LTE modelling.}

\clearpage

\end{appendix}

%
%
\end{document}